\documentclass{article}%
\usepackage{amsmath,amssymb,amsfonts,amsthm}
\usepackage{txfonts}

\usepackage{textcomp}
\usepackage{algorithm}
\usepackage{algpseudocode}
\usepackage{listings}
\usepackage{ifthen}
\usepackage[utf8]{inputenc}
\usepackage{algpseudocode}

\newtheorem{example}{Example}
\newtheorem{remark}{Remark}[section]
\newtheorem{theorem}{Theorem}
\newtheorem{definition}{Definition}
\newtheorem{corollary}{Corollary}
\newtheorem{lemma}{Lemma}
\newtheorem{proposition}{Proposition}

\newenvironment{proof_of}[1]{\noindent{\textbf{\textsf{Proof of   {#1}.}}}~}{ \hfill $\blacktriangleleft$ \vskip .2mm}

\newif\ifmai\maifalse

 \newcounter{ncomm}%

\newcommand{\BG}{$(F,G)$} 
\newcommand{\BBGG}{well-behaved} 

\newcommand{\str}[1]{\underline{{#1}}}
\newcommand{\strr}[1]{( #1 )}
\newcommand{\Pol}{{\cal P}}
\newcommand{\Mon}{{\cal M}}
\newcommand{\drifts}{{\cal D}}
\newcommand{\incond}{\rho}
\newcommand{\der}{\delta}
\newcommand{\pr}{\pi}
\newcommand{\Str}{\Sigma}
\newcommand{\conv}{\times}
\newcommand{\shu}{\otimes}
\newcommand{\had}{\odot}
\newcommand{\infil}{\uparrow}

\newcommand{\Lapl}{\mathcal{L}} 
\newcommand{\G}{\mathcal{G}} 
\newcommand{\E}{\mathcal{E}} 
\newcommand{\Lie}{\mathrm{L}} 
\newcommand{\Eqs}{\mathcal{D}}  
\newcommand{\reals}{\mathbb{R}}  
\newcommand{\C}{\mathbb{C}}  
\newcommand{\ode}{{\sc ODE}}
\newcommand{\sde}{{\sc SDE}}
\newcommand{\defi}{:=}   
\newcommand{\fact}{\textsf{fact}}
\newcommand{\convo}{\times} 
\newcommand{\K}{\mathbb{K}} 

\usepackage{chngcntr}
\usepackage{apptools}
\AtAppendix{\counterwithin{lemma}{section}}
\AtAppendix{\counterwithin{theorem}{section}}
\AtAppendix{\counterwithin{corollary}{section}}
\AtAppendix{\counterwithin{proposition}{section}}

\begin{document}

\title{Algebra and coalgebra of stream products}
\author{Michele Boreale$^1$ \qquad
Daniele Gorla$^2$\\
$^1$Universit\`a di Firenze, Italy \\
$^2$‘‘Sapienza'' Università di Roma, Italy}
\date{}

\maketitle

\begin{abstract}
We study  connections  among polynomials, differential equations and streams over a field $\K$, in terms of algebra and coalgebra. We first introduce the class of $(F,G)$-products on streams, those where the
stream derivative of a product   can be expressed as a polynomial of the streams themselves and their derivatives.
Our first result is that, for every $(F,G)$-product, there is a canonical way to construct a transition function on polynomials such that the induced  unique final coalgebra morphism from polynomials into streams is the (unique) $\K$-algebra homomorphism -- and vice-versa.
This implies one can reason algebraically on streams, via their polynomial representation. We     apply this result to obtain  an algebraic-geometric decision algorithm for polynomial stream equivalence, for an underlying generic  $(F,G)$-product. As an example of reasoning on streams, we focus   on specific products (convolution, shuffle, Hadamard) and show how to obtain closed forms of algebraic generating functions of  combinatorial sequences, as well as solutions of nonlinear ordinary  differential equations.
%
\end{abstract}

\section{Introduction}\label{sec:intro}
We   investigate   a connection  among   polynomials, differential equations and \emph{streams}, i.e., infinite sequences of elements from a set \cite{Rut03}. At a very informal level, this connection  can be expressed   by the following correspondences: polynomials =   syntax;  differential equations =  operational semantics; streams  =  abstract (denotational) semantics.
There are  two  important motivations behind this standpoint. (1) Diverse notions of \emph{product} (convolution, shuffle,...)  arise in streams, in relation to      different models -- discrete computations, combinatorial sequences, analytic functions, and more  \cite{BHPR19,Rut03}. There is also a close analogy between several forms of products and   forms  of parallelism arising in concurrency. Our aim is   to uniformly  accommodate such diverse notions,  by automatically deriving an operational semantics for polynomials that is    adequate for a given \emph{generic} stream product. (2) Once    adequate polynomial syntax and operational semantics have been obtained, one can apply powerful techniques both from  algebraic geometry (Gr\"{o}bner bases \cite{CLOS03}) and from coalgebra (coinduction \cite{Rut03})   for reasoning on streams. This includes devising algorithms for deciding stream equivalence. Again, one would like to do so in a uniform fashion w.r.t. an underlying  notion  of stream product.

Technically,   achieving these goals  amounts to defining a fully abstract semantics from polynomials to   streams,
which is essential for algebraic-geometric reasoning on streams. Moreover, one wants the resulting construction to be    as much as possible parametric with respect to the  underlying notion of stream product.

As hinted above, we will pursue these goals relying on tools from algebra and coalgebra (Section \ref{sec:background}). Indeed, it is well-known  that, when  polynomial coefficients and stream elements are drawn from  a  {field} $\K$,   both polynomials and streams  form $\K$-algebras,  i.e., rings with an additional vector space structure   over $\K$. Note that, while this algebra structure is fixed for polynomials, it varies with the underlying   product for streams. On the other hand, streams also possess a \emph{co}algebraic structure, arising from the operation of stream derivative. On the side of polynomials, it is also natural to interpret a  differential equation    $\dot x_i = p_i$ as a transition $x_i\rightarrow p_i$: thus one   expects a transition structure,  hence a coalgebra, over over polynomials as well. How to extend appropriately  transitions from individual variables $x_i$ to monomials and   polynomials, though,    depends nontrivially on the   notion of stream product one wants to model.

Our first result (Section \ref{sec:sem}) is that the above outlined goals can be achieved  for  the class of $(F,G)$-products on streams, where, basically, the
derivative of a product of two streams can be expressed as a polynomial of the streams themselves and their derivatives. One can then define
a coalgebra structure on   polynomials, depending on the given $(F,G)$-product and   differential equations, such that the unique   morphism from this coalgebra to the coalgebra of streams is also a $\K$-algebra \emph{homo}morphism (and vice-versa: every homomorphism that satisfies the given differential equations is the unique morphism).  Thus, full abstraction is achieved.

A major application of this result, which we view as our main contribution, is an algorithm   based  on  an   algebraic-geometric construction   for deciding equivalence, i.e. if two polynomials denote the same stream (Section \ref{sec:alg}). Next, focusing   on specific   $(F,G)$-products (convolution, shuffle and Hadamard; Section \ref{sec:shconv}), we show how establishing polynomial (algebraic) equalities on streams may lead to  closed forms   for generating functions of combinatorial sequences \cite{flajo}, and to solutions of nonlinear ordinary  differential equations (\ode s). In the case of convolution product, we also show that the image of the coalgebra morphism is included in the set of algebraic sequences in the sense of  \cite{flajo}.

To sum up, we make the following contributions. (1) A unifying treatment of stream products, implying that, under reasonable assumptions, coalgebra morphisms from polynomials to streams are also $\K$-algebra homomorphisms (full abstraction) -- and viceversa. (2) An algorithm for deciding polynomial stream equivalence, that relies on the full abstraction result. (3) Based on that,  methods for reasoning on generating functions and ordinary differential equations.

Due to space limitations, most proofs and additional technical material have been confined to a separate appendix.

\bigskip
\noindent
{\textbf{Related work}} Rutten's \emph{stream calculus} \cite{Rut03,Rut05}, a coinductive approach to the analysis of infinite sequences (streams), is a major source of inspiration for our work. \cite{Rut03}   studies streams, automata, languages  and formal power series  in terms of  coalgebra morphisms and bisimulation.
In close analogy with classical analysis, \cite{Rut05} presents coinductive definitions and proofs for a calculus of \emph{behavioural differential equations}, also called {\em stream differential equations} (SDEs) in later works. A number of applications  to difference equations, analytical differential equations, continued fractions  and  problems from combinatorics, are presented. Convolution and shuffle product play a central role in the stream calculus; a duality between them, mediated by a variation of Laplace transform, exists. This duality  also plays a role in our work  in relation to generating functions and solutions of \ode s (Section \ref{sec:shconv}). A coinductive treatment of analytic functions and Laplace transform is also presented by Escardo and Pavlovic \cite{Escardo}.
Basold et al. \cite{BHPR19} enrich   the stream calculus with two   types of products, Hadamard and infiltration,  and exhibit  a duality between the two,  mediated by a so-called \emph{Newton transform}.
Although these   works form  a conceptual prerequisite of our study, they do not offer a unifying  treatment of the existing disparate  notions of stream product, nor any algorithmic treatment of the induced stream equivalences.
Bonchi et al. \cite{BBBRS12} consider  an operational approach to streams and convolution product based   on weighted automata, which correspond to linear expressions. They offer an equivalence checking algorithm for such automata, and the recognized streams,    based on a  linear-algebraic construction; however, the polynomial case is not addressed.
A related work is Bonchi et al. \cite{BBHR14}, where it is shown how linear algebra and fractions can be
used to decide the equality of streams specified by linear SDEs. Here, differently from them,
we can also work with   polynomial  SDEs.

Most closely related to the present work is Hansen, Kupke and Rutten's    \cite{HKR17}. There the  authors prove that, when the SDEs defining  given operations on streams  obey  a GSOS syntactic format, then the final coalgebra morphism is also a homomorphism  from the free term algebra to the algebra (w.r.t. the given operations) of streams  \cite[Sect.8]{HKR17}.
In contrast,   we work
with
the algebra of polynomials, which  besides being  a ring  and  vector space over $\K$,   possesses  additional structure  arising from   monomials. All this   structure is essential for algebraic-geometric reasoning, and   sets our approach apart from those based on term algebras: for one thing, in term algebras there is   no obvious analog of Hilbert's basis theorem, a result  deeply related to  the well-ordering  of monomials (cf.  Dickson's lemma, \cite[Ch.2]{CLOS03}), and a crucial ingredient in  our decision algorithm.
One might  consider   more complicated GSOS frameworks      enriched with  equational theories, but even so we doubt one could naturally capture   the relevant polynomial  structure, in particular as arising from monomials. Nevertheless, a thorough  exploration of these issues  is  an interesting direction for future research.

The GSOS format has also been discussed in the framework of \emph{bialgebras} \cite[Sect.9]{HKR17}.  Bialgebras are a unified   categorical
framework  that encompass   both algebras, viewed as a   way of modeling syntax, and coalgebras, viewed as way of describing behaviours; see   \cite{Klin11} for a general introduction.  The theory of bialgebras is very abstract in spirit, and it is not immediate to pinpoint concrete relations to our results.
Furthermore, it requires a substantial  background in category theory, which we have preferred to avoid here so as to keep our approach as elementary and accessible as possible. In any case, we anticipate for bialgebras  similar difficulties to those  discussed above  for term algebras.  
For these reasons, we have preferred to leave  the exploration of   connections with bialgebras   for future work.


Somewhat related to ours is   the work of Winters on coalgebra and polynomial systems: see e.g. \cite[Ch.3]{Win14}.
Importantly, Winter considers polynomials in \emph{noncommuting} variables: under suitable assumptions, this makes his systems of equations  isomorphic  to certain context-free grammars; see also \cite{KS}. The use of noncommuting variables sets   Winter's treatment in a totally different mathematical realm, where  the algebraic geometric concepts we rely on here, like ideals and Gr\"{o}bner bases, are not applicable.

We also mention \cite{Bor19,Bor20}, that adopt  a coinductive approach to reason on polynomial  \ode s.  The ring of multivariate polynomials is employed as a syntax, with  \emph{Lie derivatives} inducing a transition structure. An algebraic-geometric algorithm to decide polynomial equivalence is presented. This algorithm as well has  inspired our decision method: in particular, as   Lie derivatives are precisely the transition structure induced in our framework by the shuffle product,  the decision algorithms of \cite{Bor19,Bor20}  are in   essence a special case of our algorithm in Section \ref{sec:alg}.
Furthermore, \cite{BorMFCS19,BorQEST20} extend   the framework of \cite{Bor19,Bor20}  to polynomial partial differential equations, which pose significative additional challenges.

Relations with work in enumerative combinatorics \cite{flajo,Stanley} are discussed in Section \ref{sec:shconv}.

\section{Background}\label{sec:background}

\subsection{Polynomials and differential equations}
Let us fix a finite, non empty set of symbols or \emph{variables}  $X = \{x_1,\ldots,x_n\}$ and a distinct variable $x\notin X$. Informally, $x$ will act as the independent variable, while $x_1,...,x_n$ will act as dependent variables, or functions,  defined by differential equations (see below).
We fix a generic field $\K$ of characteristic 0;   $\K=\reals$ and $\K=\C$ are typical choices.
We let $\Pol\defi\K[x,x_1,...,x_n]$, ranged over by $p,q,...$, be the set of polynomials with coefficients in $\K$ and indeterminates in $\{x\}\cup X$. 
We let $\Mon$, ranged over by $m,m',...$, be the set of   \emph{monomials}, that is  the free commutative monoid generated by  $\{x\}\cup X$. As usual, we shall denote polynomials as formal finite sums of distinct monomials with nonzero coefficients in $\K$: $p=\sum_{i\in I}r_i m_i$, for $r_i\in \K$. By slight abuse of notation, we shall write the zero polynomial and the empty monomial as $0$ and $1$, respectively. Over $\Pol$, one can define the usual operations of sum   $p+q$ and product $p\cdot q$, with 0 and 1 as identities,  and enjoying commutativity, associativity and distributivity, which make $\Pol$ a ring; multiplication of $p\in \Pol$  by a scalar $r\in \K$, denoted $rp$, is also defined and makes $(\Pol,+,0)$ a vector space over $\K$. Therefore,   $(\Pol, +,\times, 0, 1)$ forms a $\K$-algebra.

We shall also fix a set $\Eqs=\{\dot x_1 = p_1,...,\dot x_n = p_n\}$ of \emph{differential equations},  one for each $x_i\in X$, with $p_i\in \Pol$.  An \emph{initial condition} for $\Eqs$ is a vector $\incond=(r_1,...,r_n)\in \K^n$. The pair $(\Eqs,\incond)$ forms an
 {\em initial value problem}. The vectors $p_i$ on the right-hand side of the equations are called \emph{drifts}, and $F=(p_1,...,p_n)$ is a \emph{vector field}. Informally, each $x_i\in X$ represents a placeholder  for a function whose derivative is given by $p_i$, and whose value at the origin is $x_i(0)=r_i$.  This terminology is borrowed  from the theory of differential equations. Note, however, that depending on the semantics of polynomial product one adopts (see next section),      $\Eqs$ can be given diverse interpretations, including \emph{stream} differential equations (\sde, for convolution, see next subsection) in the sense of Rutten \cite{Rut03}, and of course \emph{ordinary} differential equations (\ode s, for shuffle).

Notationally, it will be sometimes convenient to regard $\Eqs$  and  $\incond$   as   functions $\Eqs: X \rightarrow \Pol$ and $\incond  : X \rightarrow \mathbb \K$, respectively, such that $\Eqs(x_i)=p_i$ and $\incond(x_i)=r_i$. It is also convenient to extend $\Eqs$ and $\incond$ to $x$ by letting
$\Eqs(x)=1$ and $\incond(x)=0$; note that, seen as an initial value problem, the last two equations  define the identity function. Finally, we let $x_0$ denote $x$ and, when using $\Eqs$ and $\rho$ as functions,  use $x_i$ as a metavariable on $\{x\}\cup X$: this makes $\Eqs(x_i)$ and $\incond(x_i)$  well defined for $0\leq i\leq n$.

\subsection{Streams}
\label{sec:streams}
We let  $\Str\langle \K\rangle\defi \K^\omega$, ranged over by $\sigma,\tau,...$, denote the set of \emph{streams}, that is infinite sequences  of elements from $\K$: $\sigma=(r_0,r_1,r_2,...)$ with $r_i\in \K$. Often $\K$ is understood from the context and we shall simply write $\Str$ rather than $\Str\langle \K\rangle$.
When convenient, we shall explicitly consider a stream $\sigma$ as a function from $\mathbb N$ to $\K$
and, e.g., write $\sigma(i)$ to denote the $i$-th element of $\sigma$.
By slightly overloading the notation, and when   the context is sufficient to disambiguate, the stream $(r,0,0,...)$ ($r\in \K$) will be simply denoted  by  $r$, while the stream $(0,1,0,0,...)$ will be denoted by $x$;  see \cite{Rut03} for motivations behind these notations\footnote{In particular, overloading of the symbol $x$ is
motivated by the fact that  our semantics
	of polynomials   maps the variable $x$ to  the stream $(0,1,0,0,...)$.}.
 Furthermore, a stream made up of all the same element  $r\in \K$  will be denoted as $\str r=(r,r,...)$.
One defines  the \emph{sum} of two streams $\sigma$ and $\tau$ as the stream $\sigma+\tau$ defined by: $(\sigma+\tau)(i)\defi \sigma(i)+\tau(i)$ for each $i\geq 0$, where the $+$ on the right-hand side denotes the sum in $\K$. Sum enjoys the usual commutativity and associativity properties, and has the stream $0=(0,0,...)$ as an identity.  Various forms of stream products can also be considered -- this is indeed a central theme of our paper. In particular, the {\em convolution product}
$\sigma \convo\tau $ and the {\em shuffle product} $\sigma\otimes \tau$ are defined as follows:
$(\sigma\convo \tau)(i) \defi \sum_{0\leq j\leq i}\sigma(j)\cdot\tau(i-j)$
and
$(\sigma\otimes \tau)(i) \defi  \sum_{0\leq j\leq i}{{i}\choose{j}}\sigma(j)\cdot\tau(i-j)$,
where operations on the right-hand side are carried out in $\K$ and $i\geq 0$.
The above operations enjoy alternative, easier to handle formulations based on stream differential equations -- see next subsection; there, a crucial notion will be the {\em derivative} of a stream $\sigma$, that is the stream $\sigma'$ obtained from $\sigma$ by removing its first element.

Both products are commutative,  associative,   have   $1=(1,0,0,...)$ as an identity, and distribute over $+$;   multiplication of $\sigma=(r_0,r_1,...)$  by a scalar $r\in \K$, denoted $r\sigma=(r\,r_0,r\, r_1,...)$, is also defined and makes $(\Sigma,+,0)$ a vector space over $\K$. Therefore, $(\Sigma, +,\pi,0,1)$ forms a $\K$-algebra for each of the considered product operations $\pi$.
Let us record the following useful properties for future use:
$
x \convo \sigma  = (0,  r_0, r_1,...)
$ and
$
r \,\pr\, \sigma  = (r\, r_0,r\, r_1,...) 
$, where $r\in \K$ and $\pr\in \{\convo,\otimes\}$.
In view of the second equation above, $r\,\pi\,\sigma$ coincides with
  $r\sigma$.  The first equation above leads to the so called fundamental theorem of the stream calculus, whereby for each $\sigma\in \Sigma$
\begin{equation}\label{eq:ftsc}
 \sigma  =\sigma(0)+x\times \sigma'\,.
 \end{equation}
Less commonly found forms of products, like Hadamard and Infiltration products, will be introduced in the next subsection; equations similar to \eqref{eq:ftsc} exist also for such products \cite{BHPR19,HKR17}.

\subsection{Coalgebras and bisimulation}
We quickly review some basic definitions and results about coalgebras and bisimulation; see e.g. \cite{Rut03} for a comprehensive treatment. A \emph{(stream) coalgebra with outputs in $\K$} is an automaton $C=(S,\delta,o)$, where $S$ is a nonempty set of \emph{states}, $\delta: S\rightarrow S$ is the \emph{transition} function, and $o: S\rightarrow \K$ is the \emph{output} function. A {\em bisimulation} on $C$ is a binary
relation $R \subseteq S \times S$ such that, whenever $(s,t) \in R$, then
 $o(s)=o(t)$ and $(\delta(s),\delta(t)) \in R$. As usual, there always exists a largest bisimulation on $C$, denoted $\sim$; it is the union of all bisimulations and it is an equivalence relation on $S$.
 Given two coalgebras $C_1$ and $C_2$, a \emph{coalgebra morphism} between them is a function $\mu:S_1\rightarrow S_2$ from the states of $C_1$ to the states of $C_2$ that preserves   transitions and outputs, that is (with obvious notation): $\mu(\delta_1(s))=\delta_2(\mu(s))$ and $ o_1(s)=o_2(\mu(s))$, for each $s\in S_1$.  Coalgebra morphisms preserve  bisimilarity, in the sense that $s\sim_1 t$ in $C_1$ if and only if $\mu(s)\sim_2\mu(t)$ in $C_2$. A coalgebra $C_0$ is \emph{final} in the class of coalgebras with outputs in $\K$ if, from every   coalgebra $C$ in this class, there exists a unique morphism $\mu$ from $C_0$ to $C$. In this case, $\sim_0$ in $C_0$ coincides with equality, and the following \emph{coinduction principle} holds: for every $C$ and $s\sim t$ in $C$, it holds that $\mu(s)=\mu(t)$ in  $C_0$.

The set of streams $\Sigma$ can be naturally given a stream coalgebra structure  $(\Sigma, (\cdot)',o(\cdot))$, as follows.
The {\em output} of a stream  $\sigma = (r_0,r_1,\ldots)$  is $o(\sigma)\defi r_0$ and
its {\em derivative} is $\sigma'\defi (r_1,r_2,...)$, that is $\sigma'$ is obtained  from $\sigma$ by removing its first element, that constitutes the output of $\sigma$.  In fact, this makes $\Sigma$   final   in the class of all coalgebras with outputs in $\K$ \cite{Rut03}. This also implies that one can prove equality of two streams by exhibiting an appropriate bisimulation relation relating them (coinduction).
%
%

It is sometimes convenient to consider an enhanced form of bisimulation on $\Sigma$  that relies on the notion of {\em linear closure}.\footnote{
  More general notions that we could have used here are {\em contextual closure} (see
  \cite[Thm. 2.4]{BHPR19} and works on distributive laws for bialgebras \cite{BPPR14}.
  However, the simpler notion of linear closure suffices for our purposes here.
}
Given a relation $R \subseteq \Str \times \Str$, its linear closure $\widehat R$ is
the set of pairs of the form $\left(\sum_{i=1}^n r_i\sigma_i\ ,\  \sum_{i=1}^n r_i\tau_i \right)$,
where $n \in \mathbb N$, $(\sigma_i, \tau_i) \in R$ and $r_i \in \mathbb \K$, for every $i \in
\{1,\ldots,n\}$. We say that $R$ is a {\em bisimulation up to linearity} if,
for every $(\sigma,\tau) \in R$, it holds
that $o(\sigma)=o(\tau)$ and $(\sigma',\tau') \in \widehat R$. If $R$ is a bisimulation up to linearity, then $\widehat R$ is a bisimulation \cite{Rut03};
since by definition  $R \subseteq \widehat R$, this implies that $R\subseteq\ \sim$, the bisimilarity on streams, which  coincides with equality.

A \emph{stream differential equation} (\sde) in the unknown $\sigma$  is a pair of equations of the form
$
\sigma(0)=r$
and
$\sigma'=\phi
$,
for $r\in \K$ and a stream expression $\phi$ (that can depend on $\sigma$ or its components, or even on $\sigma'$ itself). Under certain conditions on $\phi$ \cite{HKR17,Rut03},  it can be proven that there is a unique stream $\sigma$ satisfying  the above \sde. In this paper, we shall focus on the case where $\phi$ is represented by a polynomial expression -- this will be formalized in the next section. For the time being, we observe that the product operations defined in the preceding subsection enjoy a formulation   in terms of \sde s. In particular (see \cite{BHPR19,HKR17,Rut03}), for given $\sigma$ and $\tau$, their convolution and shuffle products are the unique streams satisfying the following \sde s (recall that, as a stream, $x$ denotes $(0,1,0,0,...)$):
%
\begin{align}
(\sigma \conv \tau)(0)   = \sigma(0)\cdot \tau(0) \qquad\qquad & (\sigma \conv \tau)' =  \sigma' \conv \tau  +  \sigma \conv \tau'  -  x \conv \sigma' \conv \tau' \label{ex:conv}\\
(\sigma \otimes \tau)(0)   = \sigma(0) \cdot\tau(0) \qquad\qquad & (\sigma \otimes \tau)' =  \sigma' \otimes \tau  +  \sigma \otimes \tau'\,. \label{ex:shuffle}
\end{align}
From the last equation, note the analogy between shuffle and interleaving of languages.
Moreover, the derivative of convolution product is usually defined as
$(\sigma \conv \tau)' =  \sigma' \conv \tau  +  \sigma(0) \conv \tau'$;
however, we prefer the formulation in \eqref{ex:conv} because it is symmetric.
\ifmai
where $X$ is the stream $\strr {0, 1, 0, 0, \ldots}$;
the identity for the convolution is the stream $\strr 1$.
This notion of product behaves like the concatenation (i.e., the sequential composition);
this is apparent by the alternative (but equivalent) definition for the derivative
$(\sigma \conv \tau)' = (\sigma' \conv \tau) + (\sigma(0) \conv \tau')$.
The annoying issue with this definition is that it is not symmetric.
More generally, convolution con be defined (in an almost symmetric way)
on a generic field by relying on the following expression for the derivative:
$(\sigma \conv \tau)' = (\sigma' \conv \tau(0)) + (\sigma(0) \conv \tau') + (\sigma' \conv X \conv \tau')$.
Since we are working with reals,
all definitions do coincide \cite{Rut03,BHPR19}
and hence we shall use the first one because of its perfect symmetry.
\end{example}

\begin{example}[Shuffle]\rm
\label{ex:shu}
The {\em shuffle product} '$\shu$' can be defined on streams as follows:
$$
(\sigma \shu \tau)(0) = \sigma(0) \tau(0)
\qquad\qquad
(\sigma \shu \tau)' = (\sigma' \shu \tau) + (\sigma \shu \tau')
$$

\end{example}
\fi
Two additional examples of stream products are    introduced below; see \cite{BHPR19} for the
  underlying motivations.    
%
%
The {\em Hadamard product}  $\had$  and  the {\em infiltration product} $\infil$
 can be defined by the following two \sde s.
\begin{align}
(\sigma \had \tau)(0) = \sigma(0) \tau(0)&\qquad
(\sigma \had \tau)' = \sigma' \had \tau'\\
(\sigma \infil \tau)(0) = \sigma(0) \tau(0)&\qquad
(\sigma \infil \tau)' = (\sigma' \infil \tau) + (\sigma \infil \tau') + (\sigma' \infil \tau')\,.
\end{align}
Hadamard  product $\had$ is reminiscent of synchronization in concurrency theory
and has $\str 1\defi (1,1,1,...)$ as an identity;
it is just the componentwise product of two streams, i.e.
$(\sigma \had \tau)(i) = \sigma(i) \tau(i)$, for every $i \geq 0$.
%
Infiltration product $\infil$  is again reminiscent of a notion in
 concurrency theory, namely the  fully synchronized interleaving; it has $1=(1,0,0,...)$ as an identity.



\section{(Co)algebraic semantics of polynomials and differential equations}
\label{sec:sem}
The main result of this section is that, once fixed an initial value problem $(\Eqs,\incond)$, for every product $\pi$  (with identity $1_\pr$) defined on streams and satisfying certain syntactic conditions,  one can build a coalgebra over polynomials such that the corresponding  final \emph{morphism}  into $\Sigma$  is also a $\K$-algebra \emph{homomorphism} from    $(\Pol,+,\times,0,1)$ to   $(\Sigma,+,\pi,0,1_\pi)$. In essence, the  polynomial syntax and operational semantics reflects  exactly the algebraic and coalgebraic properties of the considered $\pi$ on streams.

To make polynomials a coalgebra, we need to define the output $o:\Pol\rightarrow \K$ and transition $\delta:\Pol\rightarrow \Pol$ functions.
The definition of $o(\cdot)$ is straightforward  and only depends on the given  initial conditions $\incond$: we let $o\defi o_\incond$ be the homomorphic extension of $\incond$, seen as a function defined over $\{x\}\cup X$, to $\Pol$. Equivalently, seeing $\incond$ as a point in $\K^{n+1}$, we let $o_\incond(p)\defi p(\rho)$, that this the polynomial $p$ evaluated at the point $\incond$.
It can be easily checked that
$o_\incond(1) = 1$. 

The definition of  $\der$, on the other hand, depends on $\pi$ and is not straightforward. We will confine to products $\pi$   satisfying  \sde s of the form: $(\sigma\, \pi\, \tau)'=F(\sigma,\tau,...)$, for a given \emph{polynomial} function $F$. Then we will  require that  $\der$  on polynomials  mimics  this equation. For instance, in the case of shuffle product, we expect  that $\delta(pq)=p\delta(p)+q\delta(p)$. Therefore, our first step is to precisely define the class of products    on streams that satisfy a polynomial \sde. To this purpose, in what follows we shall consider polynomials  $G(y_1)\in \K[y_1]$ and $F(x,y_1,...,y_4)\in  \K[x,y_1,y_2,y_3,y_4]$. These  can be identified with  polynomial functions on streams: we shall write $G(\sigma_1)$, $F(x,\sigma_1,...,\sigma_4)$ for the  evaluation of $G,F$ in $(\Sigma, +,\pi,0,1_\pi)$ with specific streams $x=(0,1,0,...)$ and  $\sigma_1,...,\sigma_4$. 

\begin{definition}[\BG-product on streams]\label{def:BG}
Let $(\Sigma, +, \pi, 0, 1_\pi)$ be a $\K$-algebra, $F \in \K[x,y_1,y_2,y_3,y_4]$  and $G \in \K[y_1]$.
We say that $\pr$ is a  {\em \BG-product}
if, for each $\sigma,\tau\in \Sigma$, the following equations are satisfied:
\begin{enumerate}
\item $(\sigma\ \pr\ \tau)(0) = \sigma(0)\tau(0)$;
\item $(\sigma\ \pr\ \tau)' = F(x,\sigma, \sigma', \tau, \tau')$;

\item $1_\pr(0) = 1$ and $1_\pr' = G(1_\pr)$\,.
%
\end{enumerate}
\end{definition}

\begin{remark}
Notice that $1_\pr(0) = 1$ in Definition \ref{def:BG}(3) is a necessary condition, that follows
from Definition \ref{def:BG}(1).
Indeed, let $1_\pr(0) = r  \in \K$.
Since $1_\pr$ is the identity of $\pr$, for every $\sigma$
we must have $\sigma\, \pr\, 1_\pr =  \sigma$,
hence $(\sigma\, \pr\, 1_\pr)(0) =   \sigma(0)$. On the other hand,
by Definition \ref{def:BG}(1),   $(\sigma\, \pr\, 1_\pr)(0) = \sigma(0)\, 1_\pr(0) = \sigma(0)\, r$.
As $\sigma$ is arbitrary, we can take $\sigma(0)\neq 0$ and multiply
$\sigma(0)\, r = \sigma(0)$ by $\sigma(0)^{-1}$; this gives   $r = 1$.
However, we prefer to keep $1_\pr(0) = 1$ explicit in the definition, for the sake of clarity.
Finally, let us note that   the general theory of \sde s \cite{HKR17} ensures that conditions (1), (2), (3) in Definition \ref{def:BG} univocally define  a binary operation $\pi$ on streams,
but in general not that $\pi$ enjoys the ring axioms for product, a fact that we  must assume    from the outset.
\end{remark}

\medskip
\begin{example}
\label{ex:allprod}
For the products introduced in Section \ref{sec:background}, the pairs of  polynomials $(F,G)$ are as defined as follows.
\begin{itemize}
\item $F_\conv = y_2y_3 + y_1y_4 - xy_2y_4$. 
Note that
  $F_\conv = y_2y_3 + (y_1-xy_2)y_4$, where $ y_1-xy_2$ corresponds to $\sigma-x\convo  \sigma'=\sigma(0)$; this gives the asymmetric definition of convolution.
\item $F_\shu = y_2y_3 + y_1y_4$.
\item $F_\had = y_2y_4$.
\item $F_\infil = y_2y_3 + y_1y_4 + y_2y_4$.
\end{itemize}
The identity stream for convolution, shuffle and infiltration
is defined by
 $
1_\pr(0) = 1
$ and $
1_\pr' =   0
$,
i.e., in these cases the polynomial $G$ is $0$. For the Hadamard product, the identity is given by
$
1_\pr(0) = 1
$ and $
1_\pr' = 1_\pr
$,
i.e., the polynomial  $G$ in this case is   $y_1$.
\end{example}
\medskip

Given a \BG-product $\pi$ on streams, $\delta_\pi$ is   defined   in a straightforward manner on monomials, then extended to    polynomials by linearity. 
Below, we assume a total order on variables
 $x_0<x_1<\cdots<x_n$ and, for $m\neq 1$, let $\min(m)$ denote the smallest variable occurring in $m$ w.r.t. such a total order\footnote{
In Definition \ref{def:tr}, we are in effect totally ordering monomials by graded lexicographic order  (grlex, see \cite[Ch.1]{CLOS03}), and then proceeding by induction on this order.}.

\begin{definition}[transition function $\der_\pr$]\label{def:tr}
Let $\pr$ be a  \BG-product on streams.
We define $\delta_\pr : \Pol \rightarrow \Pol$ by induction on the size of $p\in\Pol$ as follows.
\begin{align}
\label{delta:uno}
\der_\pr(1) & =   G(1)
\vspace*{.2cm}
\\
\label{delta:due}
\der_\pr(x_i) & =   \drifts(x_i)
\vspace*{.2cm}
\\
\label{delta:tre}
\der_\pr(x_i \, m) & =   F(x,x_i, \der_\pi(x_i), m, \der_\pr(m))
\quad  \text{($m\neq 1, \;\;x_i=\min(x_i m)$)}
\\[-5pt]
\label{delta:quattro}
\der_\pr\left(\sum_{i \in I}  r_i\, m_i\right)  & =   \sum_{i \in I}  r_i\,\der_\pr(m_i)\,.
\end{align}
\end{definition}

\noindent
Returning to the products defined in Section \ref{sec:background}, we have:
$$
\begin{array}{rcl}
\der_\pi(1) & = & \left\{
\begin{array}{ll}
0 \qquad & \mbox{for $\pi\in\{\convo,\otimes,\infil\}$   (convolution, shuffle, infiltration) }
\\
1 \qquad & \mbox{for $\pi=\had$ (Hadamard)}
\end{array}
\right.
\vspace*{.3cm}
\\

\der_\pi(x_i\, m) & = & \left\{
\begin{array}{ll}
 \drifts(x_i)\cdot m + x_i\cdot\der_\pi(m) - x \cdot  \drifts(x_i)\cdot\der_\pi(m) \qquad & \mbox{for $\pi=\convo$ (convolution)}
\\
\drifts(x_i)\cdot m + x_i\cdot \der_\pi(m) \qquad & \mbox{for $\pi=\otimes$ (shuffle)}
\\
\drifts(x_i)\cdot \der_\pi(m) \qquad & \mbox{for $\pi=\had$ (Hadamard)}
\\
\drifts(x_i)\cdot m + \drifts(x_i)\cdot\der_\pi(m) + x_i\cdot\der_\pi(m) \qquad & \mbox{for $\pi=\,\infil$ (infiltration)\,.}
\end{array}
\right.

\end{array}
$$

We must now impose  certain additional sanity conditions  on $F$, to ensure that
the final coalgebra morphism induced by $\der_\pr$, as just defined, is also an algebra homomorphism.
In the rest of the paper, we will make use of the following abbreviation
$F_\pi[p;q]\defi F(x,p, \der_\pr(p),q, \der_\pr(q))$.
The necessity of the following  conditions is self-evident, if one thinks of $F_\pr[p;q]$ as
$\der_\pr(p\cdot q)$ (see Lemma \ref{lem:covid} below).

\begin{definition}[well-behaved $F$]\label{def:wb}
Let $\pr$ be a  \BG-product on streams.
We say that $\pi$ is \emph{\BBGG}
if the following equalities hold, for every $p,q \in \Pol$,
$m_1,m_2,m_i \in \Mon$, $x_i \in \{x\}\cup X$ and $r_i \in \K:$
\begin{align}
\label{F:due}
F_\pr[1;q] & =  \der_\pr(q)
\vspace*{.2cm}
\\
\label{F:tre}
F_\pr[x_im_1; m_2]  & =   F_\pr[m_1 ; x_im_2]
\vspace*{.2cm}
\\
\label{F:quattro}
F_\pr \left[\sum_{i \in I} r_i\,m_i \ ;\ q\right] & =   \sum_{i \in I} r_i\,F_\pr[m_i ; q]
\vspace*{.2cm}
\\
\label{F:cinque}
F_\pr[p;q] & =   F_\pr[q;p]\,.
\end{align}
\end{definition}

All products defined in Section \ref{sec:background} are \BBGG:
the proof of this fact, which is not entirely trivial, is reported in Appendix \ref{sec:app1}.
The following key technical result  connects morphism to homomorphism properties induced by $\pi$
and is crucial in the proof of Theorem \ref{thm:main}, that is the main result of this section;
both proofs are in  Appendix \ref{sec:app1}.

\begin{lemma}
\label{lem:covid}
Let $\pr$ be a \BBGG\ \BG-product.
Then, for every $p,q \in \Pol$, it holds that $\der_\pr(p \cdot q)  =  F_\pr[p;q]$.
\end{lemma}

\begin{theorem}
\label{thm:main}
Let $\pr$ be a \BBGG\ \BG-product.
Then  the (unique) coalgebra morphism $\mu_\pi$ from $(\Pol, \der_\pr, o_\incond)$ to
$(\Sigma, (\cdot)', o)$ is a $\K$-algebra homomorphism from $(\Pol, +, \cdot\,,0, 1)$ to
$(\Sigma,  + , \pr,  0, 1_\pr)$.
\end{theorem}

Intuitively, the proof consists in showing that $\mu_\pi$ preserves all the  operations in $\Pol$, by exhibiting in each case an appropriate bisimulation relation in $\Sigma\times \Sigma$ and then applying coinduction.  The most crucial  case is product, where one shows that the relation consisting of all pairs   $(\mu_\pi(p_1 \cdot \ldots \cdot p_k) \, ,\, \mu_\pi(p_1)\,\pr\ldots\pr\,\mu_\pi(p_k))$ ($k>0$) is a bisimulation up to linearity. Lemma  \ref{lem:covid} is used to prove that $\mu_\pi$ preserves transitions: e.g., by letting $p = p_2 \cdot \ldots \cdot p_k$, it allows one to conclude that the pair  of derivatives  $\mu_\pr( p_1\cdot p)'=\mu_\pr(F_\pr[p_1;p])$ and (slightly abusing the $F_\pr[\cdot;\cdot]$ notation) $(\mu_\pr( p_1)\,\pi\,\mu_\pr(p))'=F_\pr[\mu_\pr(p_1);\mu_\pr(p)]$ are still in relation, up to linearity.

To conclude the section, we also present a sort of converse of the previous theorem. That is,   $\mu_\pr$ is the only $\K$-algebra homomorphism that respects the
initial value problem, i.e. that satisfies $\mu_\pr(x_i)' = \mu_\pr(\drifts(x_i))$ and $\mu_\pr(x_i)(0) = \incond(x_i)$.
This is an immediate corollary of the following result and  of the uniqueness of the final coalgebra morphism.

\begin{proposition}
\label{prop:uniq}
Let $\pr$ be a \BBGG\ \BG-product and $\nu$ be a $\K$-algebra
homomorphism from $(\Pol, +, \cdot\,,0, 1)$ to
$(\Sigma,  + , \pr,  0, 1_\pr)$ that respects the initial value problem $(\drifts,\rho)$.
Then, $\nu$ is a coalgebra morphism from $(\Pol, \der_\pr, o_\incond)$ to
$(\Sigma, (\cdot)', o)$.
\end{proposition}

\newcommand{\ideal}[1]{\langle \, #1 \,\rangle}

\section{Deciding stream equality}\label{sec:alg}
One benefit of  a polynomial syntax   is the possibility of applying techniques from algebraic geometry to reason about stream equality.
We will devise an algorithm for checking whether   two given polynomials
are semantically equivalent, that is, are mapped to the same stream under $\mu_\pi$.
%
Note that, by linearity of $\mu_\pi(\cdot)$, we have that
$\mu_\pi(p) = \mu_\pi(q)$ if and only if $\mu_\pi(p)-\mu_\pi(q) = \mu_\pi(p-q) =   0$.
Therefore, checking   semantic equivalence of two polynomials reduces to the problem of checking
if a polynomial is equivalent (bisimilar) to 0.
Before introducing the actual algorithm for checking this, we quickly recall a few notions from algebraic geometry; see \cite[Ch.1--4]{CLOS03} for a comprehensive treatment.

A set of polynomials $I \subseteq \Pol$ is an {\em ideal} if
$0 \in I$ and, for all $p_1,p_2 \in I$ and $q \in \Pol$, it holds that $p_1+p_2 \in I$ and
$q \cdot p_1 \in I$.
Given a set of polynomials $S$, the {\em ideal generated by $S$} is
\begin{align*}
\ideal S&\defi  \left\{\sum_{j=1}^k  q_j \cdot p_j\ :\ k \geq 0\ \wedge\ \forall j \leq k. (q_j \in \Pol \wedge p_j \in S)\right\}\,.
\end{align*}
%
By the previous definition, we have that $\ideal{\emptyset}\defi \{0\}$. Trivially, $I=\ideal S$ is the smallest ideal containing $S$, and $S$ is called a set of \emph{generators} for $I$. It is    well-known that every ideal $I$
admits a finite set $S$ of generators (Hilbert's basis theorem). By  virtue of this result, any infinite ascending chain of ideals, $I_0\subseteq I_1\subseteq I_2\subseteq\cdots\subseteq \Pol$, stabilizes in a finite number of steps: that is, there is $k\geq 0$ s.t. $I_{k+j}=I_k$ for each $j\geq 0$ (Ascending Chain Condition, ACC). A key result due to Buchberger
is that, given a    finite $S\subseteq \Pol$, it is possible to decide whether $p\in I=\ideal S$, for any polynomial $p$. As a consequence, also ideal inclusion $I_1\subseteq I_2$ is decidable, given finite sets of generators for $I_1,I_2$. These facts are  consequences of the existence of a set  of
generators $B$ for $I$, called \emph{Gr\"{o}bner basis}, with a special property:    $p\in I$ if and only if $p\bmod B=0$, where `$\bmod \,B$' denotes the remainder of the multivariate polynomial division of $p$ by $B$. There exist   algorithms to build Gr\"{o}bner bases   which, despite their exponential worst-case complexity,  turn out to be effective in many practical cases \cite[Ch.4]{CLOS03}.

In what follows, we fix a \BBGG\ \BG-product $\pi$, and let  $\der_\pr$ and $\mu_\pr$ denote the
 associated transition  function and coalgebra morphism. Moreover, we denote by $p^{(j)}$ the $j$-th derivative of $p$, i.e. $p^{(0)} \defi  p$ and $p^{(j+1)} \defi \der_\pr(p^{(j)})$. The actual decision procedure is presented below as Algorithm \ref{alg:zerostream}. 
Intuitively, to prove that $\mu_\pr(p)=0$,
one might check if $o_\incond(p^{(j)})=0$  for every $j$, which is of course non effective. But due to ACC, 
at some point   $p^{(j)} \in \ideal{\{p^{(0)}, \ldots, p^{(j-1)}\}}$,  which implies the condition $o_\incond(p^{(j)})=0$ holds for \emph{all} $j$'s.
The correctness of this algorithm can be proven easily, under an additional mild condition on $F$: we require that $F\in \ideal{\{y_3,y_4\}}$  seen as an ideal  in $\K[x,y_1,...,y_4]$. Explicitly, $F=h_1y_3+h_2y_4$ for some $h_1,h_2\in \K[x,y_1,...,y_4]$.
The polynomials $F$ for the products in
Section \ref{sec:background} all satisfy this condition:
for example, $F_\conv = y_2y_3 + (y_1 - xy_2)y_4$.

\algnewcommand{\IfThen}[2]{
  \State \algorithmicif\ #1\ \algorithmicthen\ #2}
\algrenewcommand\algorithmicrequire{\textbf{Input:}}
\algrenewcommand\algorithmicensure{\textbf{Output:}}
\begin{algorithm}
\caption{Checking equivalence to zero}
\begin{algorithmic}[1]\label{algo:naive}
\Require{ $p \in \Pol$, a \BBGG\ \BG-product $\pi$}
\Ensure{ YES ($\mu_\pr(p)=  0$) or NO ($\mu_\pr(p)\neq   0$)}
\ForAll{$k \geq 0$}
    \IfThen{$o_\incond(p^{(k)}) \neq 0$}
    {\Return NO}  
    \IfThen{$p^{(k)} \in \ideal{\{p^{(0)}, \ldots, p^{(k-1)}\}}$}
    {\Return YES}
\EndFor
\end{algorithmic}\label{alg:zerostream}
\end{algorithm}

\begin{theorem}
\label{thm:algo}
Let $\pi$ be a \BBGG\ \BG-product, with $F\in \ideal{\{y_3,y_4\}}$.
Algorithm \ref{alg:zerostream}   terminates, and
returns YES if and only if   $\mu_\pr(p)= 0$.
\end{theorem}
\begin{proof}
Non termination  for some input polynomial   $p$ would imply that,    for all $k \geq 0$,
$p^{(k+1)} \not\in I_k\defi \ideal{\{p^{(0)}, \ldots, p^{(k)}\}}$. This in turn would imply an ever ascending chain of ideals $I_0  \varsubsetneq I_1 \varsubsetneq\cdots$, contradicting ACC.

If the algorithm returns NO, then for some $k$ we must have (recall that $\sigma^{(k)}$ stands for the $k$-th stream derivative of $\sigma$):  $o_\incond(p^{(k)})=o(\mu_\pr(p)^{(k)})= (\mu_\pr(p)^{(k)})(0)\neq 0$, thus $\mu_\pr(p) \neq   0$.

Assume now the algorithm returns YES. Then there exists   $k\geq 0$ such that
$o_\incond(p^{(j)}) = 0$, for every $0\leq j \leq k$, and
$p^{(k)} \in \ideal{\{p^{(0)}, \ldots, p^{(k-1)}\}}$. Excluding the trivial case $p=0$, we can assume $k\geq 1$.
If we prove that  $p^{(k+j)} \in \ideal{\{p^{(0)}, \ldots, p^{(k-1)}\}}$  for every $j\geq 0$,
  the thesis follows: indeed, by
$p^{(k+j)} = \sum_{i = 0}^{k-1} q_i\cdot p^{(i)}$, for some $q_i \in \Pol$, and by
$o_\incond(p^{(i)}) = 0$ for every $0\leq i \leq k-1$, it also follows $(\mu_\pr(p))(j)=(\mu_\pr(p))^{(j)}(0)=o_\incond(p^{(k+j)}) = 0$.
Now the proof that $p^{(k+j)} \in \ideal{\{p^{(0)}, \ldots, p^{(k-1)}\}}$ is by induction on $j$.
The base case ($j=0$) holds by assumption. For the induction step, let us consider
$p^{(k+j+1)}$. By definition, $p^{(k+j+1)} = \der_\pi(p^{(k+j)})$; by induction
$p^{(k+j)} = \sum_{i =0}^{k-1} q_i\cdot p^{(i)}$, for some
$q_i \in \Pol$. By \eqref{delta:quattro} and Lemma \ref{lem:covid},
$p^{(k+j+1)} = \sum_{i = 0}^{k-1} \der_\pi(q_i\cdot p^{(i)}) = \sum_{i = 0}^{k-1} F_\pr[q_i ; p^{(i)}]$.
By hypothesis $F \in \ideal{\{y_3,y_4\}}$, hence
$F_\pr[q_i;p^{(i)}] \in \ideal{\{p^{(i)},p^{(i+1)}\}}$, for every $i$, therefore
  $F_\pr[q_i;p^{(i)}] \in \ideal{\{p^{(0)}, \ldots, p^{(k-1)}\}}$, as by
hypothesis $p^{(k)} \in \ideal{\{p^{(0)}, \ldots, p^{(k-1)}\}}$.
This suffices to conclude.
\end{proof}


We first illustrate the algorithm with a simple, linear example.

\begin{example}[Fibonacci numbers]
Consider the  initial value problem $(\Eqs,\incond)$ given by the following equations.
\begin{equation}
\label{eq:fibon}
\left\{
\begin{array}{l}
\dot x_1  = x_2
\\
\dot x_2  = x_1+x_2
\end{array}
\right.
\qquad
\left\{
\begin{array}{l}
\incond(x_1) = 0
\\
\incond(x_2) = 1\,.
\end{array}
\right.
\end{equation}
Let us consider here the convolution product $\convo$. It is easily checked that $x_1$ defines the Fibonacci numbers: $\mu_\conv(x_1) = \strr{0,1,1,2,3,5,8,13,\ldots}$.
%
We want to prove the following equation:
\begin{align}
\label{eq:fib}
\mu_\conv(x_1 \cdot(1- x -x^2) ) &= \mu_\conv(x)\,.
\end{align}
Equivalently, using Algorithm \ref{alg:zerostream}, we  check that $\mu_\conv(x_1 \cdot(1- x -x^2) - x) =   0$.
Let  $p(x,x_1)\defi x_1 \cdot(1- x -x^2) - x$.
Then, an execution of Algorithm \ref{alg:zerostream} consists of the following steps.
\begin{itemize}
\item {\bf ($k$ = 0):}
 $\incond(p)=p(0,1)=0$ and
 $p^{(0)}=p(x,x_1)\notin \ideal{\emptyset}=\{0\}$.

\item {\bf ($k$ = 1):}   $p^{(1)} = x_2 \cdot (1-x-x^2) - x_1\cdot (1 + x) - x \cdot x_2 \cdot (1 + x) - 1 =
x_2 - x_1 - x_1 x - 1$. Hence, $\incond(p^{(1)}) = 1 - 1 = 0$ and $p^{(1)} \not\in \ideal{p}$.
%

\item {\bf ($k$ = 2):}    $p^{(2)} = x_1+x_2 - x_2 - (x_2x + x_1 - x x_2) = 0$. Hence,
$\incond(p^{(2)}) = 0$ and  trivially  $p^{(2)}\in \ideal{p, p^{(1)}}$.
\end{itemize}
We conclude that $\mu_\conv(p) = 0$.
\end{example}

\medskip
We now discuss a nonlinear example based   on shuffle product.

\begin{example}[double factorial of odd numbers]\label{ex:DF}
Consider the  initial value problem $(\Eqs,\incond)$ given by the following equation.
\begin{equation}
\label{eq:df}
\begin{array}{l}
\dot y  = y^3
\end{array}
\qquad
\begin{array}{l}
\incond(y) = 1\,.
\end{array}
\end{equation}
Let us consider here the shuffle product $\otimes$. It is easily checked that $\mu_\otimes(y) = \strr{1, 1, 3, 15, 105, 945, 10395, 135135,\ldots}$, the sequence of double factorials of odd numbers (sequence  A001147 in \cite{OEIS}).
%
We want to check the following equation
\begin{align}
\label{eq:dfeq}
\mu_\otimes(y^2(x - 1/2) + 1/2) &= 0\,.
\end{align}
using Algorithm \ref{alg:zerostream}. 
Let  $q(x,y)\defi y^2(x - 1/2) + 1/2$.
An execution of Algorithm \ref{alg:zerostream} consists of the following steps.
\begin{itemize}
\item {\bf ($k$ = 0):}
 $\incond(q)=q(0,1)=0$ and
 $q^{(0)}=q(x,y)\notin \ideal{\emptyset}=\{0\}$.

\item {\bf ($k$ = 1):}   $q^{(1)} = 2y^4 x - y^4 + y^2 =
2y^2 q$,  hence     $q^{(1)} \in \ideal{q}$.
%
\end{itemize}
We conclude that $\mu_\otimes(q) = 0$.
\end{example}

\begin{remark}\label{rem:gf}
Note that we can define the generating function associated to Fibonacci numbers, that is the function  $g(z)$ whose Taylor series expansion  is  $ \sum_{j \geq 0} f_j z^j$ (where  $f_j$ are the Fibonacci numbers); such a generating function is
\begin{align}
\label{eq:fib1}
g(z) &= \frac z {1-z-z^2}\,.
\end{align}
Now, from \cite{BHPR19} it is known  that the convolution product admits
an inverse of a given stream $\sigma$ whenever $\sigma(0) \neq 0$. Thus,
from \eqref{eq:fib} we obtain
$
\mu_\conv(x_1) = \mu_\conv(x) \conv (\mu_\conv(1-x-x^2))^{-1} =
\frac {\mu_\conv(x)} {1- \mu_\conv(x) - \mu_\conv(x)^2}
$,
where we use the usual notation $\frac \sigma \tau $ to denote $\sigma \conv \tau^{-1}$.
This   equation for $\mu_\conv(x_1)$ is structurally identical to   \eqref{eq:fib1}: this is of course no coincidence, as algebraic identities on streams correspond  exactly to algebraic identities on generating functions.  This will be made precise in the next section -- see in particular Proposition \ref{prop:algconv}.

Similarly, the equivalence $\mu_\otimes(p)=0$ obtained for the double factorial equations, when solved algebraically for $x_1$ yields the exponential generating function for A001147, that is $g(z)=\sqrt{1/(1-2z)}$: see Example \ref{ex:DF2} in Subsection \ref{sub:ode}.
\end{remark}

We finally point out that Algorithm \ref{alg:zerostream} can be easily modified to actually \emph{find} all polynomials $p$,  up to a prescribed degree,  s.t. $\mu_\pi(p)=0$, along the lines of a similar procedure in \cite{Bor20}. Indeed, we actually found the polynomials in both examples above using this modified algorithm\footnote{Python code, with instructions and examples, available at {\tt https://local.disia.unifi.it/boreale/papers/streams.py}.}.


\section{Shuffle,  convolution and generating functions}\label{sec:shconv}
\newcommand{\Q}{\mathbb{Q}}
We study  the relation of the shuffle and convolution products, and of the corresponding morphisms,    with algebraic  sequences arising in enumerative combinatorics \cite{flajo,Stanley}, and with solutions of ordinary  differential equations; Hadamard product plays also a role in connecting the other two products. Our aim here is not to prove any new  identity, but rather to relate our framework with certain well established notions and results in these fields. In particular,  we will argue that our results can be useful for combinatorial reasoning on sequences and   \ode s: this means   chiefly  finding  generating functions of sequences,  \ode\ solutions, and/or establishing nontrivial  relations among them.

\subsection{Generating functions}
For a stream $\sigma=(r_0,r_1,...,r_j,...)$, we let the \emph{ordinary generating function} \cite{flajo,Stanley} of  $\sigma$ in the variable $z$   be the power series
$\G[\sigma](z)\defi \sum_{j\geq 0} r_j z^j\,.$
We shall normally understand $\G[\sigma](z)$   as a  {formal} power series, which is just another   convenient, functional    notation for the stream $\sigma$. When $\K=\reals$ or $\K=\C$,   it is sometimes convenient to   consider $z$ as a real or complex variable\footnote{For example, the study of the generating function in a complex analytic sense, in particular of its poles, provides detailed information on the asymptotic growth of the elements of $\sigma$; see \cite{flajo}.}:
 in this case,
 $\G[\sigma](z)$ defines a (real or complex) analytic function around $0$,
 provided that its radius of convergence is positive. In fact, we shall see that, when $\sigma=\mu_\conv(p)$, then $\G[\sigma](z)$ is analytic. We denote by $\G^{-1}[g(z)]$ the inverse transformation, mapping a power series $g(z)=\sum_{j\geq 0} r_j z^j$   back to $\sigma=(r_0,r_1,...)$. More precisely, for any (formal or analytic) power series $g(z)$ around the origin, $\G^{-1}[g(z)]$ can be obtained by taking the Taylor coefficients of $g(z)$:
\begin{align}\label{eq:invG}
\G^{-1}[g(z)]&= \left(\frac{g^{(0)}(0)}{0!}, \frac{g^{(1)}(0)}{1!}, \frac{g^{(2)}(0)}{2!},...\right)
\end{align}
where $g^{(j)}(z)$ denotes the $j$-th derivative of $g(z)$, in either formal or analytic sense. With the same convention on $z$, we let  the \emph{exponential} generating function of $\sigma$ to be the Taylor series
$\E[\sigma](z)\defi \sum_{j\geq 0} \frac{r_j}{j!}z^j\,.$
Again, $\E^{-1}[g(z)]$ denotes the inverse transformation, mapping a (formal or analytic) power series   $g(z)$ to the stream of its derivatives evaluated at 0:
\begin{align}\label{eq:invE}
\E^{-1}[g(z)]&= ( {g^{(0)}(0)} ,  {g^{(1)}(0)} ,  {g^{(2)}(0)} ,...)\,.
\end{align}
Letting $\fact\defi (0!,1!,2!,...)$ and $\exp(z) \defi \sum_{j\geq 0}\frac {z^j} {j!}$,  the relation between $\G$ and $\E$ can be written as follows, where  the Hadamard product on power series is defined as $(\sum_j a_jz^j)\had (\sum_j b_jz^j)\defi \sum_j (a_jb_j)z^j $  as expected:
\begin{align}
\E[\sigma](z) =\exp(z)\had \G[\sigma](z)\label{eq:EG2}
\\
\E^{-1}[g(z)] =\fact\had \G^{-1}[g(z)]\label{eq:GE}\,.
\end{align}
Again, for $\sigma=\mu_\otimes(p)$, we will see that $\E[\sigma](z)$ is analytic. The maps $\G[\cdot]$ and $\E[\cdot]$  act as $\K$-algebra
 homomorphisms between streams and   functions. In particular,
  products of streams is transformed into   product of functions\footnote{When interpreted in a purely formal sense, hence   in terms of streams: the  equation for $\G$ just defines an alternative notation  for convolution product; the  equation for $\E$ reduces to \eqref{eq:transform}.}, that is \cite{flajo,Stanley}:
$$
\G[\sigma\times \tau](z)=\G[\sigma](z)\cdot\G[\tau](z)
\qquad\qquad
\E[\sigma\otimes \tau](z)=\E[\sigma](z)\cdot\E[\tau](z)\,.
$$
These relations allow  one to transform algebraic equations on streams into algebraic equations on generating
functions. One reason to  perform   this transformation is that, if a closed expression for
 the generating function can be  found via analytic  manipulations,  the actual stream  can be recovered by applying the inverse transforms \eqref{eq:invG} and \eqref{eq:invE} -- that is essentially via   Taylor expansion.

\subsection{Algebraic streams}
In what follows, we let $p$ range over $\Pol=\K[x,x_1,...,x_n]$  and $q=q(x,y)$ over $\K[x,y]$, while $g(z)$ still denotes a formal power series or analytic function at the origin.

\begin{definition}[algebraic  streams, \cite{flajo}]\label{def:alghol} A function $g(z)$ is
\emph{algebraic}
if there is a nonzero polynomial $q(x,y)$   such that $q(z,g(z))$ is identically 0. In this case, $g(z)$ is called a \emph{branch} of $q(x,y)$.
A stream $\sigma$ is algebraic if $\G[\sigma](z)$ is algebraic.
\end{definition}


If the degree of $q(x,y)$ in $y$ is $k$, then $q(x,y)$  has at most $k$  branches. For example, $q(x,y)=y^2+x-1$ has  two distinct branches, that is algebraic functions: $g(z)=\pm\sqrt{1-z}$. When the coefficients of $q$ are drawn from a subfield of $\C$, then it can be shown   that the corresponding branches are also complex analytic (hence real analytic when restricted to $\reals$); see \cite{Ahlfors}.
Our starting point in the study of the connections between  coalgebra morphisms and algebraic streams   is the following simple result.

\begin{proposition}\label{prop:algconv}
Let $p\in \Pol$  and  $\sigma=\mu_\convo(p)$. Suppose there is a polynomial $q(x,y)\neq 0$ such that
  $\mu_{\convo}(q(x,p))=0$. Then $\G[\sigma](z)$ is a branch of $q$.
%
The corresponding statement for $\mu_\otimes(\cdot)$ and $\E[\cdot]$ is also true.
\end{proposition}
\begin{proof}
We only check  the case of the convolution product.
We know that $\mu_\convo(\cdot)$   is a $\K$-algebra 
homomorphism from polynomials to streams. Moreover, $\mu_\convo(x)=x=(0,1,0,...)$.
We have already noted that $\G[\cdot]$,   as a mapping from streams to   power series, is a $\K$-algebra
 homomorphism as well. Moreover, $\G[x](z)=z$, the identity function. These facts imply that $\G[\mu_\convo(q(x,p))](z)=q(z,\G[\sigma](z) )=0$. The last equation precisely says that $\G[\sigma](z)$ is a
 branch  of $q(x,y)$.
\end{proof}

Pragmatically, the above result    implies  that, if one     proves  a nontrivial polynomial equation $q(x,\sigma)=0$ for $\sigma=\mu_\pi(p)$  ($\pi\in\{\convo,\otimes\}$), e.g. by using the algorithm in the previous section, then one can recover $\sigma$ by Taylor expansion of one of the branches of $q$; see Example \ref{ex:catalan} below.

In the case of  the convolution product $\convo$, the result  also implies that,  under the given hypotheses, $\sigma$ is algebraic.
In fact, something  more general can be said. Let the considered  system of  differential  equations and initial conditions be    $\Eqs=\{{\dot x}_1 = p_1,...,{\dot x}_n=p_n\}$ and $ \incond  =(r_1,...,r_n)\in \K^n$, respectively; let $\sigma_i\defi \mu_\convo(x_i)$ for $i=1,...,n$. As a consequence of \eqref{eq:ftsc}, it is easy to check that the streams $\sigma_i$, hence the corresponding    generating functions $\G[\sigma_1](z),...,\G[\sigma_n](z)$, satisfy the following   system  of polynomial equations in the variables $x_1,...,x_n$:
\begin{align}\label{eq:polysystem}
x_1 = r_1 + x   p_1 &\qquad\cdots\qquad
x_n  = r_n + x  p_n \,.
\end{align}
In the terminology of Kuich and Salomaa \cite[Ch.14]{KS}, \eqref{eq:polysystem} is a \emph{weakly strict} polynomial system (in the single letter alphabet $\{x\}$). They prove that  there is a unique tuple of formal power series that solves this system, which therefore coincides with $(\sigma_1,...,\sigma_n)$.
Moreover, by invoking  elimination theory, Kuich and Salomaa prove  that, for each $i=1,...,n$, \eqref{eq:polysystem} implies a nontrivial polynomial equation $q(x,x_i)=0$ for the variable $x_i$: see \cite[Ch.16, Cor.16.11]{KS}, which covers the case $\K=\Q$. We sum up the above discussion in the following.

\begin{corollary}[algebraicity of  $\mu_\convo$]\label{th:alg}
Suppose that $\K=\Q$. 
Then, for each $p\in \Pol$, $\mu_\convo(p)$ is an algebraic stream  in the sense of Definition \ref{def:alghol}.
\end{corollary}

\newcommand{\xx}{\mathrm{\mathbf{x}}}
When $\K=\Q$,  the above result implies that $\G[\mu_\conv(p)](z)$ is analytic. At present we do not know if the  converse of this corollary is true, i.e. if all algebraic functions are expressible via polynomial \sde.

\begin{example}[Catalan   numbers]\label{ex:catalan}
Let $\K=\reals$. Consider the differential equation in one dependent variable (here $y=x_1$)
\begin{align}\label{eq:cat}
\dot y &=y^2
\end{align}
with the initial condition $y(0)=1$. Let us analyse this equation from the point of view of convolution product.
%
By  \eqref{eq:ftsc},  we have $\mu_\convo(y)=\mu_\convo(y)(0)+x\times (\mu_\convo(y))'=1+x\times\mu_\convo(\delta_\convo (y))=1+x\times\mu_\convo( y^2)=1+x\times\mu_\convo( y)^2$. Let $\sigma=\mu_\convo(y)$, this leads to the polynomial equation $q(x,\sigma )=0$, where $q(x,y)=y-xy-y^2-1$. Solving for $y$ as a function of $x$ (and renaming $x$ to $z$), we obtain two branches, $y(z)=(1\pm \sqrt{1-4 z})/2 z$. By Proposition \ref{prop:algconv},  $\sigma$ must be the series of Taylor coefficients of one or the other of these two functions.  One checks that  the stream obtained using the minus sign solves the equation:
\begin{align*}
\sigma\,=\,\G^{-1}\left[\frac{1 - \sqrt{1-4 z}}{2 z}\right]&=(1, 1, 2, 5, 14, 42, 132,...)\,.
\end{align*}
These are the Catalan numbers, sequence A000108 in \cite{OEIS}.
\end{example}
%
%
%

\subsection{Solutions of \ode s}\label{sub:ode}

The shuffle product $\otimes$ provides a connection between streams and differential equations.  A recurrent motif here is that streams and their generating functions can be used to reason on solutions of \ode s -- and the other way around. In what follows, solutions might be considered in both formal and analytic sense.

When applied to $\otimes$, Proposition \ref{prop:algconv}     may   help  one to recover   closed forms for algebraic solutions of a \ode\ system, in case they exist. This is entailed by Corollary \ref{cor:algsol} below.
In the rest of the section, we let $\xx(z)=(x_1(z),...,x_n(z))$ denote a  solution around 0 of   $\Eqs=\{{\dot x}_1 = p_1,...,{\dot x}_n=p_n\}$, considered as a   system of \ode s,  with the given initial conditions $\xx(0) \defi \incond\in \K^n$.  In particular, note that, when $\K=\reals$,    a solution always exists, is unique and analytic (Picard-Lindel\"{o}f theorem). For  $p(x,x_1,...,x_n)\in \Pol$, we let $p(z,\xx(z))$ denote the composition of $p$ as a function with $(z,\xx(z))$;  in turn, $p(z,\xx(z))$ is   a formal power series or analytic function around the origin.  The following proposition provides a link between solutions of \ode s and shuffle product and the induced morphism, via exponential generating functions. The essential point here is that $\delta_\otimes$ coincides with Lie derivative; a proof is in the Appendix \ref{sec:app2}.

\begin{proposition}\label{prop:shLap}
$p(z,\xx(z))=\E[\mu_\otimes(p)](z)$.
\end{proposition}

When $\K=\reals$, the above result implies that $\E[\mu_\otimes(p)](z)$ is always real analytic.

\begin{corollary}[algebraic solutions of \ode s]\label{cor:algsol}
Suppose that,  for some nonzero $q=q(x,y)$, we have $\mu_\otimes(q(x,p))=0$. Then $p(z,\xx(z))$ is a branch of $q(x,y)$.
\end{corollary}
\begin{proof}
By Proposition \ref{prop:algconv}, 
we deduce that $\E[\mu_\otimes(p)](z)$ is a branch of $q(x,y)$. But, by Proposition \ref{prop:shLap}, $p(z,\xx(z))=\E[\mu_\otimes(p)](z)$. 
\end{proof}

A discussion on the relation of $\mu_\otimes$ with   algebraic and other classes of streams is deferred  to the end of the section. We illustrate now the above results with a simple example.

\begin{example}[factorial numbers and the solution of $\dot y = y^2$]\label{ex:factorial}
Consider again the equation $\dot y  =y^2$  with $y(0)=1$ of Example \ref{ex:catalan}. This time
we analyse this equation from the point of view of   shuffle product.
Let $\sigma=\mu_\otimes(y)$. Consider the polynomial $q=q(x,y)\defi yx-y+1$. One checks that $q\sim 0$  in the coalgebra over   $\Pol$ induced by $\delta_\otimes$: to see this, one applies the algorithm in Section \ref{sec:alg}, noting that $o(q)=q(0,1)=0$ and that $\delta_\otimes(q)=yq\in \langle q\rangle$. This implies $\mu_\otimes(q(x,y))=0$, hence, according to   Proposition \ref{prop:algconv}, $\E[\sigma](z)$ is a branch of $q(x,y)$.
Now $q(x,y)$ defines a unique branch, $y(z)=\frac 1{1-z}$. Then using also \eqref{eq:GE}:
\begin{align*}
\sigma\,=\,\E^{-1}\left[\frac{1}{1-z}\right]&=\fact\had \G^{-1}\left[\frac{1}{1-z}\right]
                                             = \fact\had (1,1,1,...)
                                             = (0!,1!,2!,...)\,.
\end{align*}
Finally, by Corollary  \ref{cor:algsol}, the solution of \eqref{eq:cat} as an \ode\ with the initial condition $y(0)=1$  is the unique branch of $q$, that is $y(z)=\frac 1 {1-z}$.
\end{example}

\begin{example}[double factorials, again]\label{ex:DF2}
Consider again the equation $\dot y  =y^3$  with $y(0)=1$ of Example \ref{ex:DF}, and the equivalence
$\mu_\otimes(q)=0$, for $q(x,y)\defi y^2(x - 1/2) + 1/2$, we proved there. Let $\sigma=\mu_\otimes(y)$.  According to   Proposition \ref{prop:algconv}, the exponential generating function   $\E[\sigma](z)$ is a branch of $q(x,y)$. Now $q(x,y)$ has two branches, which are obtained by solving for $y$ the corresponding quadratic equation. Of these,   $y(z)= \sqrt{1/(1-2z)}$ solves the \ode\ and, by Proposition \ref{prop:shLap}, is the exponential generating function of $\sigma$.
\end{example}

Let us also point out an interesting interplay between $\convo$ and $\otimes$,  that may ease compositional reasoning on streams. Depending on the equations at hand,  the  convolution   of two streams might be more easily understood and described than their shuffle product; or a stream can be better understood in terms of the solution  of an \ode. 
The following equality, that can be readily checked, allows one to transform convolution into stream product, and back. We let $\fact^{-1}\defi(1/0!, 1/1!,...,1/j!,...)$.
\begin{align}\label{eq:transform}
\fact^{-1}\had(\sigma\otimes \tau)=(\fact^{-1}\had \sigma)\convo(\fact^{-1}\had \tau)\,.
\end{align}
We illustrate this idea with a simple example.

\begin{example}[harmonic numbers]\label{ex:sh2conv}
Consider the system of two equations $\dot y = y^2$, $\dot w=y$ with initial conditions $y(0)=1$ and $w(0)=0$. 
We want to analyze this system in terms of $\otimes$. In     Example \ref{ex:factorial}, we have   seen that $y(z)=\frac 1{1-z}$ and that $\sigma\defi \mu_\otimes(y)=\fact$. We can obtain $\mu_\otimes(w)$ via Proposition \ref{prop:shLap} and \eqref{eq:GE}, after solving the second \ode: $w(z)=\int_0^z y(u)du=\ln(\frac 1{1-z})$, hence $\tau\defi \mu_\otimes(w)=\E^{-1}[w(z)]=\fact\had (0, 1, 1/2,...,1/j,...)$. To understand what $\mu_\otimes(yw)=\mu_\otimes(y)\otimes\mu_\otimes(w)$ represents, it is convenient to switch to the convolution product, by applying \eqref{eq:transform}.
We have
\begin{align*}
\fact^{-1}\had \mu_\otimes(yw)&=\fact^{-1}\had (\sigma\otimes\tau)
 =(\fact^{-1}\had \sigma)\convo(\fact^{-1}\had \tau)\\
& = (1,1,1,...)\convo (0, 1, 1/2,...,1/j,...)
 = (0,1,3/2,...,\sum_{i=1}^j \frac 1 i,...)
\end{align*}
which is the sequence $\alpha=(h_0,h_1,...)$ of  the  {harmonic} numbers. Therefore $\mu_\otimes(yw)=\fact\had\alpha$, and $y(z)w(z)=\frac 1{1-z}\ln(\frac 1 {1-z}) =\E[\fact\had\alpha](z)=\sum_{j\geq 0}h_j z^j=\G[\alpha](z)$ is the ordinary generating function of the harmonic numbers.
\end{example}

\begin{remark}\label{rem:shuffle}
One would like to prove  for $\mu_\otimes$ a    result  analogous to Corollary \ref{th:alg}. In this respect, let us first note that $\mu_\otimes(p)$ need not be algebraic: as we have seen in Example \ref{ex:catalan},   $\mu_\otimes(y)=\fact = (0!,1!,2!,...)$, which is  {not} an algebraic stream -- cf. \cite{Stanley}, or simply note that $\G[\fact](z)$ is not analytic. The next natural candidate class  to consider for inclusion is that of streams with a \emph{holonomic} (a.k.a. \emph{$D$-finite}) ordinary generating function \cite{Stanley}: that is, a function $y(z)$ satisfying a linear differential equation with polynomial coefficients  in $z$. This class includes strictly algebraic streams, but   $\mu_\otimes(p)$ need not be holonomic either. To see this, consider the single \ode\ $\dot f = 1+f^2$  with $f(0)=0$, which defines the trigonometric tangent function: $f(z)=\tan(z)$. It is known that $\sigma=\G^{-1}[\tan(z)]$ is not holonomic, see e.g. \cite[Ch.1]{Mallinger}. It is also known that $\fact$ is holonomic, and that the class of holonomic functions is closed under the Hadamard product \cite{Stanley}. Now, from Proposition \ref{prop:shLap} and \eqref{eq:GE}, we have that: $\mu_\otimes(f)=\E^{-1}[\tan(z)]=\fact\had \sigma$. This equality implies that $\mu_\otimes(f)$  is not holonomic, because otherwise $\sigma$ would be as well.  At present,  we also ignore if  algebraic and/or holonomic streams are included in streams obtainable via $\mu_\otimes$.
\end{remark}

\subsection{Linear \ode s and Laplace transform}\label{sub:linodes}
Another example of interplay between   shuffle and convolution products arises in connection with the  solutions of linear \ode s, that is when the drifts $p_i$ ($i=1,...,n$)   are linear functions of $x_1,...,x_n$.
%
Most of the times,  Corollary \ref{cor:algsol} will not apply to   {linear} \ode s. In particular, when $\K=\C$, a nontrivial linear system  will only   posses  transcendental solutions -- exponential or trigonometric.  In handling the linear case, however,   it is much simpler to  rely    on the  {convolution} product $\times$.

First, note that when only linear equations $\Eqs$ and expressions $f$ are involved,
 one has $\mu_\otimes(f)=\mu_\conv(f)$: this can be proven   either directly, or by noting that the streams $\sigma_i\defi \mu_\otimes(x_i)$ ($i=1,...,n$) as well are the (unique) solution of the  system \eqref{eq:polysystem} in the \emph{convolution} ring $(\Sigma,+,\convo,0,1)$ --  a consequence of the linearity of $\mu_\otimes(\cdot)$ and of  \eqref{eq:ftsc}. Moreover, the system \eqref{eq:polysystem}  can be written in the form $\xx^T=\incond^T+xA\xx^T$, for $\xx=(x_1,...,x_n)$ and a suitable matrix $A\in \K^{n\times n}$.   The vector of ordinary generating functions $H(z)\defi (\G[\sigma_1](z),...,\G[\sigma_n](z))$  as well  satisfies this linear   system, and can be given an explicit formula, as follows. Below, we consider  $n\times n$ matrices of power series;    power series in $z$    with scalar matrices as coefficients are interpreted element-wise, that is as defining matrices of power series. With these conventions, we can   solve the   system \eqref{eq:polysystem} for $H(z)$, and after some linear algebra  arrive at
\begin{align}\label{eq:Gz}
H(z)^T&=   (I-zA)^{-1}\incond^T\,=\;(I+zA+z^2A^2+\cdots +z^jA^j+\cdots)\incond^T\,.
\end{align}
Note that, in a formal sense,   $(I-zA)^{-1}$ always exists (moreover, if   $z$ is interpreted as a real or complex variable,   $H(z)$   is analytic around the origin). By Proposition \ref{prop:shLap}, the  solution vector of the given initial value problem  is $\xx(z)= (\E[\sigma_1](z),...,\E[\sigma_n](z))$. Recalling the relation \eqref{eq:EG2} between ordinary and exponential generating functions,  the matrix exponential notation $\exp(zA)=\sum_{j\geq 0}\frac{A^j}{j!}z^j$, and letting $\had$   distribute  componentwise on vectors of functions, we have
\begin{align*}
\xx(z)^T = \exp(z)\had H(z)^T
    = \exp(z)\had  (I-zA)^{-1}\rho^T
    = \exp(zA)\rho^T
\end{align*}
which is a well known formula for the solutions of   linear systems of \ode s.

\begin{example}[a linear system of \ode s]
As a simple example of application of the above method, consider the linear system $\dot x_1=-x_2, \dot x_2=x_1$,  with $x_1(0)=1, x_2(0)=0$, corresponding to  $A=    \left[\begin{smallmatrix} -1&0\\ 0&1 \end{smallmatrix}\right]$ and   $\incond=(1,0)$. Then $H(z)^T=(I-zA)^{-1}\incond^T= \left(  \frac {1}{{z}^{2}+1} ,  \frac {z}{{z}^{2}+1}  \right)^T$. Taylor expanding, we obtain e.g. $H_1(z)= 1-{z}^{2}+{z}^{4}+\cdots $, hence $x_1(z)=\exp(z)\had H_1(z)=1-{\frac {1}{2}}{z}^{2}+{\frac {1}{24}}{z}^{4}+ \cdots$, which one recognizes as $\cos(z)$. Similarly, $x_2(z)=\sin(z)$. The   calculation of the explicit formula for $\exp(z)\had  (I-zA)^{-1}\rho^T$ can be automatized   via the partial fraction decomposition of $(I-zA)^{-1}\rho^T$ \cite{flajo}.
\end{example}

\newcommand{\Lap}{\mathcal{L}}
The analogy of the above solution procedure with that based on Laplace transform should be evident. Note in particular that the function $H(z)$ in \eqref{eq:Gz} is similar to, \emph{but not the same   as}, the Laplace transform of $\xx(z)$.  Indeed,  applying  the true Laplace transform to  the linear \ode\ system  $\dot\xx^T =A\xx^T$   with initial condition $\xx(0)=\incond$,  we have $\Lap[\dot\xx^T](z)=z\Lap[\xx^T](z)-\xx^T(0)=A\Lap[\xx^T](z)$, hence
$\Lap[\xx^T ](z)=(zI-A)^{-1}\rho^T$
(note the different position of the $z$ variable). Likewise, $\exp(z)\had H(z)$ is similar to, but not the same as,   taking the inverse Laplace transform of $\Lap[\xx^T](z)$.

It is remarkable that $H(z)$ allows one to solve a system  of linear \ode s exactly and rigorously,   without any reference to  notions  from classical calculus. This is certainly true for solutions  in a formal sense. Analyticity   then follows by either appealing to classical results (Picard-Lindel\"{o}f theorem), or directly noting that $H(z)$, hence $\exp(z)\had H(z)$, is analytic at the origin by construction.
In the context of streams and coalgebra, similarities and differences between ordinary generating functions and the Laplace transform have been noted elsewhere \cite{Escardo,Rut05}.



\section{Conclusion}\label{sec:concl}
We have studied connections between polynomials, differential equations and streams, in terms of algebra and coalgebra. Our main result shows that, given any stream product that satisfies certain reasonable assumptions, there is a way to define a transition function on polynomials such that the induced unique coalgebra morphism into streams is a  $\K$-algebra homomorphism -- and vice-versa. We have   applied this result to the design of a decision algorithm for polynomial stream equivalence, and to reasoning on generating functions and ordinary differential equations.

As for future work, it would be interesting to see whether we can define new notions of products
that respect the format we devised in this paper. Somewhat orthogonal to this,
  the relation of our framework with bialgebras \cite{Klin11} deserves further investigation. Finally, in the field of  nonlinear dynamical  systems \cite{Khalil}, convolution of discrete sequences arises as a means to describe the composition of distinct signals or subsystems (e.g.,  a plant and a controller); we would like to understand if our approach can be useful to reason on such systems as well.


\appendix
\section{Proofs and additional technical material}\label{sec:app}

\subsection{Proofs of Section \ref{sec:sem}}
\label{sec:app1}
Here, we check that  the shuffle product is \BBGG\ (the other products are dealt with similarly).
To this aim, we need a preliminary Lemma.

\begin{lemma}
\label{lemma:Fder}
For every $m_1,m_2 \in \Mon$,  it holds that $F_\shu[m_1;m_2] = \der_\shu(m_1m_2)$.
\end{lemma}
\begin{proof}{Lemma \ref{lemma:Fder}}
By induction on $m_1$. The base case is simple, once we notice that a monomial $m$ corresponds to the
polynomial mapping 1 to $m$ and 0 to every other monomial; then, by \eqref{F:due},
$F_\shu[1;m_2] = \der_\shu(m_2) = \der_\shu(1\, m_2)$. For the inductive step,
let $m_1$ be different from $1$ and let $x_i$ be the variable with lower index in $m_1m_2$;
assume that $x_i$ comes from $m_2$ (the proof can be done similarly if it comes from $m_1$),
and let $m_2'$ be $m_2$ without one occurrence of $x_i$. Then:
$$
\begin{array}{lll}
F_\shu[m_1;m_2]
& = \der_\shu(m_1) \cdot m_2 + m_1\cdot \der_\shu(x_i m_2')
\vspace*{.2cm}
\\
& = \der_\shu(m_1) \cdot m_2 + m_1\cdot F_\shu[x_i; m_2']
\vspace*{.2cm}
\\
&= \der_\shu(m_1) \cdot x_i m_2' + m_1\cdot (\der_\shu(x_i)\cdot m_2' + x_i \cdot \der_\shu(m_2'))
\vspace*{.2cm}
\\
&= \der_\shu(m_1)\cdot x_i m_2'  + \der_\shu(x_i)\cdot m_1m_2' + x_i m_1\cdot \der_\shu(m_2')
\vspace*{.2cm}
\\
&= \der_\shu(x_i)\cdot m_1m_2' + x_i \cdot (\der_\shu(m_1)\cdot m_2' + m_1\cdot \der_\shu(m_2'))
\vspace*{.2cm}
\\
&= \der_\shu(x_i)\cdot m_1m_2' + x_i \cdot F_\shu[m_1;m_2']
\vspace*{.2cm}
\\
&= \der_\shu(x_i)\cdot m_1m_2' + x_i \cdot \der_\shu(m_1m_2')
\vspace*{.2cm}
\\
&= F_\shu[x_i;m_1m_2']
\vspace*{.2cm}
\\
&= \der_\shu(x_i\ m_1m_2')
\vspace*{.2cm}
\\
&=\der_\shu(m_1m_2)
\end{array}
$$
where the first, the third, the sixth and the eighth steps are by definition of $F_\shu$,
the second and the ninth ones are by \eqref{delta:tre},
the seventh one is by induction, and the remaining ones by usual operations on
polynomials/monomials.
\end{proof}

\noindent
Recall that $F_\shu = y_2y_3 + y_1y_4$, see Example \ref{ex:allprod}. Then:
\begin{itemize}
\item property \eqref{F:due} is satisfied, since $F_\shu[1;q] = 0\cdot q + 1\cdot \der_\shu(q) = \der_\shu(q)$;
    \item property \eqref{F:tre} follows from Lemma \ref{lemma:Fder}.
Indeed, $F_\shu[x_im_1;m_2] = \der_\shu(x_im_1m_2)  = \der_\shu(m_1x_im_2) = F_\shu[m_1;x_im_2]$;

\item property \eqref{F:quattro} holds. Indeed:
$$
\begin{array}{lll}
F_\shu \left[\sum_{i \in I} r_i\,m_i \ ;\ q\right]
& = \der_\shu\left(\sum_{i \in I} r_i\,m_i \right)\cdot q + \left(\sum_{i \in I} r_i\,m_i \right) \cdot \der_\shu(q)
\vspace*{.2cm}
\\
& = \left(\sum_{i \in I} r_i\,\der_\shu(m_i) \right)\cdot q + \left(\sum_{i \in I} r_i\,m_i \right) \cdot \der_\shu(q)
\vspace*{.2cm}
\\
& = \sum_{i \in I} r_i\,(\der_\shu(m_i) \cdot q) + \sum_{i \in I} r_i\,(m_i \cdot \der_\shu(q))
\vspace*{.2cm}
\\
& = \sum_{i \in I} r_i\,(\der_\shu(m_i) \cdot q + m_i \cdot \der_\shu(q))
\vspace*{.2cm}
\\
& = \sum_{i \in I} r_i\,F_\shu[m_i ; q]
\end{array}
$$
where the first and last equality hold by definition of $F_\shu$, the second one by \eqref{delta:quattro}, and
the remaining ones by distributivity of sum over the product in polynomials;

\item property \eqref{F:cinque} trivially holds.
\end{itemize}

\bigskip
\bigskip

\begin{proof_of}{Lemma \ref{lem:covid}}
We proceed by cases on $p$.
\begin{description}
\item[$p = 0$.] \ \vspace*{-.4cm}
$$
\begin{array}{lll}
\der_\pr(p \cdot q) & = \der_\pr(0)  & \mbox{by def. of product in $\Pol$}
\vspace*{.2cm}
\\
& = 0 & \mbox{by \eqref{delta:quattro}}
\vspace*{.2cm}
\\
& = F_\pr[0 ; q] & \mbox{by \eqref{F:quattro}\,.}
\end{array}
$$

\item[$p=m \in \Mon$.]
If $m = 1$, we trivially conclude by \eqref{F:due}, since $p\cdot q = q$.
Otherwise, we first consider the case in which $q=m' \in \Mon$.
Let $x_i$ be the variable with smallest index
in $m \cdot m'$ and $m''$ be $m \cdot m'$ with  one occurrence of $x_i$ removed. Then
$\delta_\pr(m \cdot m') = \der_\pr(x_i \cdot m'')$,
by commutativity and associativity in $\Mon$. Now
    \begin{itemize}
    \item if $m'' = 1$, then $m' = 1$ (i.e., that $q = 1$) and $m = x_i$; then
    $\der_\pr(x_i \cdot m'') = \der_\pr(x_i) = F_\pr[m'';x_i] = F_\pr[m;m']$,
    by identity of the product, \eqref{F:due} and \eqref{F:cinque};
    \item otherwise, $\der_\pr(x_i \cdot m'') = F_\pr[x_i ; m''] = F_\pr[m ; m']$, by
    \eqref{delta:tre} and \eqref{F:tre} (applied $|m|-1$ times).
    \end{itemize}
Now, let $q=\sum_{j \in J} r_j\,m_j$, for $|J| > 0$. We have
$$
\begin{array}{lll}
\der_\pr(p \cdot q) & = \der_\pr(\sum_{j \in J} r_j\,(m \cdot m_j)) &
\mbox{by distributivity in $\Pol$}
\vspace*{.2cm}
\\
& = \sum_{j \in J} r_j\,\der_\pr(m \cdot m_j) &
\mbox{by \eqref{delta:quattro}}
\vspace*{.2cm}
\\
& = \sum_{j \in J} r_j\,F_\pr[m ; m_j] &
\mbox{by the previous discussion for $q$ a monomial}
\vspace*{.2cm}
\\
& = \sum_{j \in J} r_j\,F_\pr[m_j ; m] &
\mbox{by \eqref{F:cinque}}
\vspace*{.2cm}
\\
& = F_\pr[q;m] &
\mbox{by \eqref{F:quattro}}
\vspace*{.2cm}
\\
& = F_\pr[p;q] &
\mbox{by \eqref{F:cinque}\,.}
\end{array}
$$

\item[$p=\sum_{i \in I} r_i\,m_i$, for $|I| > 0$.]
$$
\begin{array}{lll}
\der_\pr(p\cdot q) & = \der_\pr(\sum_{i \in I} r_i\,(m_i \cdot q)) &
\mbox{by distributivity in $\Pol$}
\vspace*{.2cm}
\\
& = \sum_{i \in I} r_i\,\der_\pr(m_i \cdot q) &
\mbox{by \eqref{delta:quattro}}
\vspace*{.2cm}
\\
& = \sum_{i \in I} r_i\,F_\pr[m_i ; q] &
\mbox{by the previous discussion for $p$ a monomial}
\vspace*{.2cm}
\\
& = F_\pr[p ; q] &
\mbox{by \eqref{F:quattro}\,.}
\end{array}
\vspace*{-.6cm}
$$
\end{description}
\end{proof_of}

\bigskip
\bigskip

In the following   proofs, we will use the following notation.
Given a polynomial substitution (a  map  from variables to polynomials) $\zeta$,
and a monomial $m=x_{i_1}\cdots x_{i_k}$, we let  $m\zeta$ denote the polynomial $\zeta(x_{i_1})\cdot \ldots \cdot \zeta(x_{i_k})$.
Similarly, given a stream substitution (a  map  from variables to streams)  $\xi$,
  we let $m\xi$ denote the stream $\xi(x_{i_1})\,\pr \cdots \pr\, \xi(x_{i_k})$.

\bigskip
\begin{proof_of}{Theorem \ref{thm:main}}
We prove that $\mu=\mu_\pi$ preserves the ring operations and their identities,  as well as
 multiplication by a scalar.  It suffices to prove the following three statements, as the preservation of 0 follows from the first point (see below).
\begin{enumerate}
\item $\mu(r_1p+r_2q) = r_1\mu(p)+r_2\mu(q)$.
We prove that
$$
R = \{ (\mu(r_1p+r_2q) \, , \, r_1\mu(p)+r_2\mu(q))\ :\ p,q \in \Pol, r_1,r_2 \in \K\}
$$
is a bisimulation. Pick up any $(\mu(r_1p+r_2q) \, , \, r_1\mu(p)+r_2\mu(q)) \in R$,
we need to prove two conditions.
    \begin{enumerate}
    \item $\mu(r_1p+r_2q)(0) = (r_1\mu(p)+r_2\mu(q))(0)$: since $\mu$ is a coalgebra morphism and by
    definitions, $\mu(r_1p+r_2q)(0) = o_\incond(r_1p+r_2q) = r_1\,o_\incond(p) + r_2\,o_\incond(q) =
    r_1\mu(p)(0)+r_2\mu(q)(0) = (r_1\mu(p)+r_2\mu(q))(0)$.
    \item $\mu(r_1p+r_2q)' = (r_1\mu(p)+r_2\mu(q))'$: since $\mu$ is a coalgebra morphism, by \eqref{delta:quattro}
    and by definitions, $\mu(r_1p+r_2q)' = \mu(\der_\pr(r_1p+r_2q)) = \mu(r_1\der_\pr(p)+r_2\der_\pr(q))$, whereas
    by definition and by the fact that $\mu$ is a coalgebra morphism,
    $(r_1\mu(p)+r_2\mu(q))' = r_1\mu(p)' + r_2\mu(q)' =  r_1\mu(\der_\pr(p)) + r_2\mu(\der_\pr(q))$.
    This suffices to conclude, since $(\mu(r_1\der_\pr(p)+r_2\der_\pr(q))\ , \ r_1\mu(\der_\pr(p)) + r_2\mu(\der_\pr(q))) \in R$.
    \end{enumerate}

\item $\mu(1) = 1_\pr$. It suffices to prove that
$$
R = \{\ (\mu(1) \, ,\, 1_\pr)\ \}
$$
is a bisimulation up to linearity.
To this aim, we need to check two conditions.
    \begin{enumerate}
    \item $\mu(1)(0) = 1_\pr(0)$. Since $\mu$ is a coalgebra morphism,
    by definition of $1$, and by Def.\,\ref{def:BG}(3),
    we have that $\mu(1)(0) = o_\incond(1) = 1 = 1_\pr(0)$.

    \item $\mu(1)' = 1_\pr'$. Since $G$ is a polynomial
    in the variable $y_1$ (i.e. $G = \sum_{i\in I} r_i\,m_i$, where the $m_i$'s are monomials
    in $y_1$), we have that
    $$
    \begin{array}{lll}
    \mu(1)' & = \mu(\der_\pr(1)) & \mbox{$\mu$ is a coalgebra morphism}
    \vspace*{.2cm}
    \\
    & = \mu(G(1)) & \mbox{by \eqref{delta:quattro} and \eqref{delta:due}}
    \vspace*{.2cm}
    \\
    & =\ \sum_{i \in I} r_i\,\mu(m_i\zeta) & \mbox{point 1 of this proof}
    \vspace*{.2cm}
    \\
    & =\ \sum_{i \in I} r_i\,\mu(1)
    \end{array}
    $$
    where $\zeta$ is the substitution that maps $y_1$ to $1$, hence
    all monomials $m_i$ evaluated under $\zeta$
    yield $1$, which justifies the last step above.
    By Definition \ref{def:BG}(3) and definition of $G$, we have that
    $ 1_\pr' = G(1_\pr) = \sum_{i \in I} r_i\,(m_i\xi)  = \sum_{i \in I} r_i\,1_\pr$
    where $\xi$ is the substitution that maps $y_1$ to $1_\pr$,  hence
    all monomials $m_i$ evaluated under $\xi$ yield  $1_\pr$.
    This suffices to conclude up to linearity.
    \end{enumerate}

\item $\mu(p \cdot q) = \mu(p)\ \pr\ \mu(q)$. To prove this fact, let us consider the relation
$$
R = \{ (\mu(p_1 \cdot \ldots \cdot p_k) \, ,\, \mu(p_1)\,\pr\ldots\pr\,\mu(p_k)) \ : \ p_i \in \Pol, k > 0\ \}
$$
and prove that it is a bisimulation up to linearity.
Let us consider any $(\sigma,\tau)=(\mu(p_1 \cdot \ldots \cdot p_k) \, ,\, \mu(p_1)\,\pr\ldots\pr\,\mu(p_k)) \in R$.
We will prove that (a) $o(\sigma)=\sigma(0)=\tau(0)=o(\tau)$ and (b) $(\sigma',\tau')\in \widehat R$ (the linear closure of $R$).
The   case  $k=1$  is trivial, so assume $k>1$.
Let     $q=p_2 \cdot \ldots \cdot p_k$.
We check conditions (a) and (b) defined above.
    \begin{enumerate}
    \item[(a)] $\mu(p_1 \cdot q)(0) = (\mu(p_1)\,\pr\,\mu(q))(0)$:
    $$
    \begin{array}{lll}
    \mu(p_1 \cdot q)(0) & = o_\incond(p_1 \cdot q) & \mbox{since $\mu$ is a coalgebra morphism}
    \vspace*{.2cm}
    \\
    & = o_\incond(p_1)o_\incond(q) & \mbox{by def. of } o_\incond(\cdot)
    \vspace*{.2cm}
    \\
    &  = \mu(p_1)(0)\mu(q)(0) & \mbox{since $\mu$ is a coalgebra morphism}
    \vspace*{.2cm}
    \\
    & = (\mu(p_1)\,\pr\,\mu(q))(0) & \mbox{by Definition \ref{def:BG}(1).}
    \end{array}
    $$
    \item[(b)] $\mu(p_1 \cdot q)' \;\widehat R\; (\mu(p_1)\,\pr\,\mu(q))'$. By assumption $F$ is a polynomial
    in the variables $x,y_1,\ldots,y_4$, i.e. $F = \sum_{i\in I} r_i\,m_i$, where $r_i\in \K$ and the $m_i$'s are monomials
    in these variables. Let $\zeta:  \{x,y_1,\ldots,y_4\} \rightarrow \Pol$
    and $\xi : \{x,y_1,\ldots,y_4\} \rightarrow \Sigma$ be the substitutions defined as follows.
    $$
    \begin{array}{llll}
      \zeta:    &  x \ \mapsto x
      &
    \xi:  & x\ \mapsto \mu(x)
    \\
      &  y_1 \mapsto p_1
      &
      & y_1 \mapsto \mu(p_1)
    \\
    & y_2 \mapsto \der_\pr(p_1) \qquad\qquad
    &
    & y_2 \mapsto \mu(\der_\pr(p_1))
    \\
    & y_3 \mapsto q
    &
    & y_3 \mapsto \mu(q)
    \\
    & y_4 \mapsto \der_\pr(q) \qquad
    &
    & y_4 \mapsto \mu(\der_\pr(q))\,.
    \end{array}
    $$
    Then  we have
    $$
    \begin{array}{lll}
    \mu(p_1 \cdot q)' & =\ \mu(\der_\pr(p_1 \cdot q)) & \mbox{$\mu$ is a coalgebra morphism}
    \vspace*{.2cm}
    \\
    & =\ \mu(F_\pr[p_1 ; q]) & \mbox{Lemma \ref{lem:covid}}
    \vspace*{.2cm}
    \\
    & =\ \mu\left(\sum_{i \in I} r_i\,(m_i\zeta) \right)& \mbox{def.\,of $F[\ \cdot\ ;\ \cdot\ ]$}
    \vspace*{.2cm}
    \\
    & =\ \sum_{i \in I} r_i\,\mu(m_i\zeta) & \mbox{part 1 of this proof}
    \end{array}
    $$
    and
    $$
    \begin{array}{lll}
    (\mu(p_1)\, \pr\, \mu(q))' & =\ F(\mu(p_1), \mu(p_1)', \mu(q), \mu(q)' )
    & \mbox{Def. \ref{def:BG}(2)}
    \vspace*{.2cm}
    \\
    & =\ F( \mu(p_1), \mu(\der_\pr(p_1)), \mu(q), \mu(\der_\pr(q)) )
    & \mbox{$\mu$ coalgebra morphism}
    \vspace*{.2cm}
    \\
    & =\ \sum_{i \in I} r_i\,(m_i\xi)    & \mbox{def.\,of $F$.}
    \end{array}
    $$
    Now by definition   $(\mu(m_i\zeta)\, ,\, m_i\xi) \in R$  for every $i \in I$. Thus  we can conclude
    up to linearity.
    \end{enumerate}
\end{enumerate}
Finally, concerning the identity of sum, by part 1 we have  $\mu(0)=\mu(p-p)=\mu(p)-\mu(p)=0$.
\end{proof_of}

\bigskip
\bigskip

\begin{proof_of}{Proposition \ref{prop:uniq}}
We have to prove two properties for $\nu$,
for every $p \in \Pol$.
\begin{description}
\item[1. $o_\incond(p) = \nu(p)(0)$.] To this aim, let us first prove that
$o_\incond(m) = \nu(m)(0)$, for every $m \in \Mon$; the proof is by induction on $m$.
\begin{itemize}
\item Base ($m = 1$). $o_\incond(1) = 1 = 1_\pr(0) = \nu(1)(0)$, where
the first equality holds by definition, the second one by Def.\,\ref{def:BG}(3),
and the last one by homomorphism of $\nu$.
\item Induction ($m = x_i m_1$).
$\nu(x_i m_1)(0) = (\nu(x_i)\ \pr\ \nu(m_1))(0) = (\nu(x_i)(0))(\nu(m_1)(0)) =
\incond(x_i) o_\incond(m_1) = o_\incond(m)$, where the first equality holds by homomorphism,
the second one by Def.\,\ref{def:BG}(1), the third one since $\nu$ respects $\incond$
and by induction, and the last one by definition.
\end{itemize}
Now, let $p = \sum_i r_im_i$. Then, $o_\incond(p) = \sum_i r_i o_\incond(m_i) =
\sum_i r_i (\nu(m_i)(0)) = \nu(p)(0)$, where the first equality holds by definition of $o_\incond$,
the second one by this claim for monomials, and the third one by algebra homomorphism of $\nu$.

\item[2. $\nu(\der_\pi(p)) = \nu(p)'$.]
We proceed by cases on $p$.
\begin{description}
\item[$p = 0$.] In this case,
$\nu(\der_\pr(0)) = \nu(0) =  0 = \nu(0)'$, where the first equality holds by
\eqref{delta:quattro}, the second one by homomorphism, and the third one by definition of $ 0$ and
homomorphism of $\nu$.

\item[$p = m  \in \Mon$.]
The proof is by induction on $m$.
\begin{itemize}
\item $m = 1$. In this case, let $G$ be $\sum_i r_im_i$, where the $m_i$'s are monomials in $y_1$.
Then, by letting $\zeta$ be the substitution that maps $y_1$ to $1$, we have that:
$$
\begin{array}{lll}
\nu(\der_\pr(1)) & = \nu(G(1)) &
\mbox{by \eqref{delta:uno}}
\vspace*{.2cm}
\\
& = \sum_i r_i\,\nu(m_i\zeta) &
\mbox{by def. of $G$ and homomorphism of $\nu$}
\vspace*{.2cm}
\\
& = \sum_i r_i\,\nu(1) &
\mbox{since every $m_i$ is a monomial only in $z$}
\vspace*{.2cm}
\\
& = \sum_i r_i\,1_\pr &
\mbox{by homomorphism of $\nu$}
\vspace*{.2cm}
\\
& = G(1_\pr) &
\mbox{by def. of $G$ and the fact that $m_i$'s are monomials only in $y_1$}
\vspace*{.2cm}
\\
& = 1_\pr' &
\mbox{by Def.\,\ref{def:BG}(3)}
\vspace*{.2cm}
\\
& = \nu(1)' &
\mbox{by homomorphism of $\nu$.}
\end{array}
$$

\item $m = x_i$. In this case, $\nu(\der_\pr(x_i)) = \nu(\drifts(x_i)) = \nu(x_i)'$, where the
first equality holds by \eqref{delta:due}, and the second one since $\nu$ respects $\drifts$.

\item $m = x_i \bar m$, for $\bar m \neq 1$ and $x_i$ the variable with smallest index in $m$.
Let $F$ be $\sum_j r_j m_j$, where the $m_j$'s
are monomials in the variables $x,y_1,\ldots,y_4$. For every such a monomial $m_j$, let us
denote by $e_{j,k}$ the exponent of variable $y_k$ in $m_j$, for $k=0,...4$, where we let $y_0=x$; i.e. $m_j = x^{e_{j,0}}y_1^{e_{j,1}}y_2^{e_{j,2}}y_3^{e_{j,3}}y_4^{e_{j,4}}$. Furthermore, let us define the substitutions $\zeta$
and $\xi$ as follows:
    $$
    \begin{array}{llll}
    \zeta\ : & x\ \mapsto x
    &
    \xi\ : & x\ \mapsto \nu(x)\\
       & y_1 \mapsto x_i
    &
      & y_1 \mapsto \nu(x_i)
    \\
    & y_2 \mapsto \drifts(x_i) \qquad\qquad
    &
    & y_2 \mapsto \nu(x_i)'
    \\
    & y_3 \mapsto \bar m
    &
    & y_3 \mapsto \nu(\bar m)
    \\
    & y_4 \mapsto \der_\pr(\bar m) \qquad
    &
    & y_4 \mapsto \nu(\bar m)'\,.
    \end{array}
    $$
By assumption  on $\nu$ (homomorphism and respects   $(\drifts,\incond)$) and by induction on $\bar m$, which   is smaller than $m$,  we have the following, where $(\nu(u))^e$ stands for $\nu(u)\,\pi\,\cdots\,\pi\,\nu(u)$ ($e$ times):
\begin{eqnarray}
\nonumber
\nu(m_j\zeta)
& = & (\nu(x))^{e_{j,0}}\ \pr\  (\nu(x_i))^{e_{j,1}}\ \pr\ (\nu(\drifts(x_i)))^{e_{j,2}}\ \pr\ (\nu(\bar m))^{e_{j,3}}
\ \pr\ (\nu(\der_\pr(\bar m)))^{e_{j,4}}
\vspace*{.2cm}
\\
\nonumber
& = & (\nu(x))^{e_{j,0}}\ \pr\ (\nu(x_i))^{e_{j,1}}\ \pr\ (\nu(x_i)')^{e_{j,2}}\ \pr\ (\nu(\bar m))^{e_{j,3}}
\ \pr\ (\nu(\bar m)')^{e_{j,4}}
\vspace*{.2cm}
\\
& = & m_j\xi\,.
\label{eq:monSubs}
\end{eqnarray}
Putting all together,  we   obtain the desired result for $m$:
$$
\begin{array}{lll}
\nu(\der_\pr(x_i\bar m)) & = \nu(F(x_i,\drifts(x_i), \bar m, \der_\pr(\bar m))) &
\mbox{by \eqref{delta:tre} and \eqref{delta:due}}
\vspace*{.2cm}
\\
& = \nu(\sum_{j} r_j m_j\zeta) &
\mbox{by definition of $F$ and $\zeta$}
\vspace*{.2cm}
\\
& = \sum_{j} r_j \, \nu(m_j\zeta) &
\mbox{by homomorphism of $\nu$}
\vspace*{.2cm}
\\
& = \sum_{j} r_j \, m_j\xi &
\mbox{by \eqref{eq:monSubs}}
\vspace*{.2cm}
\\
& = F(\nu(x_i), \nu(x_i)', \nu(\bar m), \nu(\bar m)') &
\mbox{by definition of $F$ and $\xi$}
\vspace*{.2cm}
\\
& = (\nu(x_i)\ \pr\ \nu(\bar m))' &
\mbox{by Def.\,\ref{def:BG}(2)}
\vspace*{.2cm}
\\
& = \nu(x_i\bar m)' &
\mbox{by homomorphism of $\nu$.}
\end{array}
$$
\end{itemize}

\item[$p=\sum_{i \in I} r_i\,m_i$, for $|I| > 0$.]
$$
\begin{array}{lll}
\nu(\der_\pr(p)) & = \nu(\sum_{i \in I} r_i\,\der_\pr(m_i)) &
\mbox{by \eqref{delta:quattro}}
\vspace*{.2cm}
\\
& = \sum_{i \in I} r_i\,\nu(\der_\pr(m_i)) &
\mbox{since $\nu$ is an algebra homomorphism}
\vspace*{.2cm}
\\
& = \sum_{i \in I} r_i\,(\nu(m_i)') &
\mbox{by the case for $p$ a monomial}
\vspace*{.2cm}
\\
& = \nu(p)' &
\mbox{by linearity of stream derivatives.}
\end{array}
\vspace*{-.6cm}
$$
\end{description}
\end{description}
\end{proof_of}

\subsection{Proofs of Section \ref{sec:shconv}}
\label{sec:app2}

Using the notation of Section \ref{sec:shconv}, we let $p$ range over $\Pol=\K[x,x_1,...,x_n]$, let $\xx(z)=(x_1(z),...,x_n(z))$ denote a solution of the system of \ode s with the initial condition $\xx(0)=\incond=(r_1,...,r_n)\in \K^n$, and let
 $p(z,\xx(z))$ be the composition of $p$ with $(z,\xx(z))$. In turn, $p(z,\xx(z))$ is an analytic function/formal power series centered at the origin $z=0$:
\begin{align}\label{eq:qtaylor}
p(z,\xx(z))&=\sum_{j\geq 0} \frac 1{j!} p^{(j)}(0)z^j
\end{align}
where
$p^{(j)}\defi \frac{\mathrm{d}^j}{\mathrm{d} z^j} p(z,\xx(z))$. We consider the \emph{Lie derivative} of the polynomial $p(x,x_1,...,x_n)$ w.r.t. the extended vector field $\overline F\defi (p_0,p_1,...,p_n)$, where $p_0\defi 1$, is (recall that by convention $x_0=x$)
\begin{align*}
\Lie(p)&\defi \langle \nabla p\,,\,\overline F\rangle\, =\, \sum_{i=0}^n \frac{\partial p}{\partial x_i} \cdot  p_i
\end{align*}
which is in turn a polynomial in $\Pol=\K[x,x_1,...,x_n]$.  The $j$-th Lie derivative $\Lie^{(j)}(p)$, $j\geq 1$, is defined by induction as $\Lie^{(j+1)}(p)=\Lie(\Lie^{(j)}(p))$.
The following relation between Lie derivatives and ordinary derivatives of $p(z,\xx(z))$ is well-known (see e.g. \cite[p.3,\ eqs. (4,5)]{Bor20}), for each $j\geq 0$, letting $\overline{\xx}(0)\defi(0,\xx(0))$
\begin{align}\label{eq:Lie}
p^{(j)}(0)\,\defi\,\frac{\mathrm{d}^j }{\mathrm{d} z^j } p(z,\xx(z))_{|z=0} & = (\Lie^{(j)}(p))(\overline\xx(0))\,.
\end{align}

The proof of the following lemma is straightforward and omitted.

\begin{lemma}\label{lemma:Lie}
For each     $p$,  $\delta_\otimes(p)=\Lie(p)$.
\end{lemma}

\begin{proof_of}{Proposition \ref{prop:shLap}}
First, Lemma \ref{lemma:Lie} extends  to $j\geq 0$, that is:
  $\delta_\otimes^{(j)}(p)=L^{(j)}(p)$. Exploiting also  the properties of coalgebra morphisms and \eqref{eq:Lie}, we have,  for each $j\geq 0$:
\begin{align*}
(\mu_\otimes(p))(j) & = (\mu_\otimes(p))^{(j)}(0)\\
                  & = (\mu_\otimes(\delta^{(j)}_\otimes(p)))(0)\\
                  & = (\mu_\otimes(\Lie^{(j)}(p)))(0)\\
                  & = o(\mu_\otimes(\Lie^{(j)}(p)))\\
                  & = o(\Lie^{(j)}(p))\\
                  & = (\Lie^{(j)}(p))(\overline\xx(0))\\
                  & = p^{(j)}(0)\,.
\end{align*}
In other words, $\mu_\otimes(p)=(p^{(0)}(0),p^{(1)}(0),...,p^{(j)}(0),...)$. Taking $\E[\cdot]$ on both sides of this equality  and considering \eqref{eq:qtaylor}, we get the thesis. 
\end{proof_of}


\end{document}